\documentclass[12pt]{article}
\usepackage{fullpage,citesort,epsfig,psfrag,
graphics,amsbsy,amssymb,
}
\usepackage{caption}
\usepackage{subcaption}
\newcommand{\beq}{\begin{equation}}
\newcommand{\eeq}{\end{equation}}
\newcommand{\bea}{\begin{eqnarray}}
\newcommand{\eea}{\end{eqnarray}}
\newcommand{\bfs}{\boldsymbol}

\newcommand{\be}{\begin{equation}}
\newcommand{\ee}{\end{equation}}
\newcommand{\bq}{\begin{eqnarray}}
\newcommand{\eq}{\end{eqnarray}}
\newcommand{\ket}[1]{|#1\rangle}

\def\math{\mathsurround=0pt }
\def\leftrightarrowfill{$\math \mathord\leftarrow \mkern-6mu
 \cleaders\hbox{$\mkern-2mu \mathord- \mkern-2mu$}\hfill
 \mkern-6mu \mathord\rightarrow$}
\def\overleftrightarrow#1{\vbox{\ialign{##\crcr
     \leftrightarrowfill\crcr\noalign{\kern-1pt\nointerlineskip}
     $\hfil\displaystyle{#1}\hfil$\crcr}}}

\newcommand{\VEV}[1]{\langle#1\rangle}

   \let\i=\iota 
\let\l=\lambda

 \def\bd{\begin{document}} \def\ed{\end{document}}
\def\ds{\documentstyle} \let\fr=\frac \let\bl=\bigl \let\br=\bigr
\let\Br=\Bigr \let\Bl=\Bigl
\let\bm=\bibitem
\let\na=\nabla
\let\pa=\partial \let\ov=\overline
\def\ft#1#2{{\textstyle{{\scriptstyle #1}\over {\scriptstyle #2}}}}
\def\fft#1#2{{#1 \over #2}}
\def\vp{\varphi}
\def\sst#1{{\scriptscriptstyle #1}}
\def\oneone{\rlap 1\mkern4mu{\rm l}}
\def\td{\tilde}
\def\wtd{\widetilde}
\def\dalemb#1#2{{\vbox{\hrule height .#2pt
        \hbox{\vrule width.#2pt height#1pt \kern#1pt
                \vrule width.#2pt}
        \hrule height.#2pt}}}
\def\square{\mathord{\dalemb{6.8}{7}\hbox{\hskip1pt}}}
\def\wtd{\widetilde}
\def\R{\rlap{\rm I}\mkern3mu{\rm R}}
\def\im{{\rm i}}
\def\tilg{\tilde{g}}
\def\tilF{\tilde{F}}
\def\tilA{\tilde{A}}
\def\varf{\varphi}
\def\tilf{\tilde{\phi}}
\def\tilh{\tilde{h}}
\def\rme{{\rm e}}
\def\ep{\epsilon}
\def\0{{(0)}}
\def\9{{(9)}}
\def\8{{(8)}}
\def\7{{(7)}}
\def\6{{(6)}}
\def\5{{(5)}}
\def\4{{(4)}}
\def\3{{(3)}}
\def\2{{(2)}}
\def\1{{(1)}}
\newcommand{\trace}{{\rm Tr}}
\newcommand{\ub}{\overline{U}}
\newcommand{\vb}{\overline{V}}
\newcommand{\uh}{\widehat{U}}
\newcommand{\vh}{\widehat{V}}
\newcommand{\ubh}{\overline{\widehat{U}}}
\newcommand{\vbh}{\overline{\widehat{V}}}
\newcommand{\lb}{\bar{\l}}
\newcommand{\Fb}{\overline{F}}
\newcommand{\Fh}{\widehat{F}}
\newcommand{\Fbh}{\overline{\widehat{F}}}
\newcommand{\Ab}{\overline{A}}
\newcommand{\Ah}{\widehat{A}}
\newcommand{\Abh}{\overline{\widehat{A}}}
\newcommand{\Gb}{\overline{G}}
\newcommand{\Gh}{\widehat{G}}
\newcommand{\Gbh}{\overline{\widehat{G}}}
\newcommand{\Pb}{\overline{P}}
\newcommand{\Ph}{\widehat{P}}
\newcommand{\Pbh}{\overline{\widehat{P}}}
\newcommand{\Qb}{\overline{Q}}
\newcommand{\Qh}{\widehat{Q}}
\newcommand{\Qbh}{\overline{\widehat{Q}}}
\newcommand{\Bb}{\overline{B}}
\newcommand{\Bh}{\widehat{B}}
\newcommand{\Bbh}{\overline{\widehat{B}}}
\newcommand{\fhns}{\hat{F}^{\rm (NS)}}
\newcommand{\fhrr}{\hat{F}^{\rm (RR)}}
\newcommand{\ahns}{\hat{A}^{\rm (NS)}}
\newcommand{\ahrr}{\hat{A}^{\rm (RR)}}
\newcommand{\hhrr}{\hat{H}^{\rm (RR)}}
\newcommand{\hchi}{\hat{\chi}}
\newcommand{\hphi}{\hat{\phi}}
\newcommand{\htau}{\hat{\tau}}
\newcommand{\cG}{{\cal G}}
\newcommand{\cGb}{\overline{{\cal G}}}
\newcommand{\cH}{{\cal H}}
\newcommand{\cP}{{\cal P}}
\newcommand{\cPb}{\overline{{\cal P}}}
\newcommand{\cQ}{{\cal Q}}
\newcommand{\cQb}{\overline{{\cal Q}}}
\newcommand{\cM}{{\cal M}}
\newcommand{\cN}{{\cal N}}
\newcommand{\cO}{{\cal O}}
\newcommand{\cD}{{\cal D}}
\newcommand{\cL}{{\cal L}}

\newcommand{\vpp}{\mbox{$\langle{\scriptstyle++}\rangle$}}
\newcommand{\vmp}{\mbox{$\langle{\scriptstyle-+}\rangle$}}
\newcommand{\vppp}{\mbox{$\langle{\scriptstyle+++}\rangle$}}
\newcommand{\vmpp}{\mbox{$\langle{\scriptstyle-++}\rangle$}}
\newcommand{\vpmp}{\mbox{$\langle{\scriptstyle+-+}\rangle$}}

\begin{document}
\setlength{\captionmargin}{36pt}
\begin{titlepage}
\begin{flushright}
\phantom{.}
\end{flushright}

\vskip 3cm
\begin{center}
\begin{large}
{\bf Open String Self-energy on the Lightcone Worldsheet Lattice}
\end{large}

\vskip 2cm
{\large
Georgios Papathanasiou\footnote{E-mail  address: {\tt georgios@ufl.edu}} and Charles B. Thorn\footnote{E-mail  address: {\tt thorn@phys.ufl.edu}}
}
\vskip0.20cm
{\it Institute for Fundamental Theory\\
Department of Physics, University of Florida,
Gainesville FL 32611}


\vskip 1.0cm
\end{center}

\begin{abstract}
\noindent We continue our study of open string perturbation theory
on the lightcone worldsheet lattice, which is an $M\times N$
rectangular grid. Here $M$ is the number of $P^+$ units and
$N$ is the number of $ix^+$ units. We extend our previous
analysis to the bosonic open string one planar
loop self-energy. We find that, when all open string
coordinates satisfy Neumann conditions,  the ultraviolet worldsheet
divergences associated with the closed string tachyon
and boundary effects can be cancelled  by renormalization of bulk ($AM^1$) and
boundary ($BM^0$) worldsheet ``cosmological constants''.  The bulk divergence
for the open string matches that for the closed string.
The open string tachyon mass shift displays
the dilaton logarithmic divergence with the correct coefficient
for its consistent absorption by renormalization of the string tension.
The ultraviolet contribution to the open string gluon mass shift vanishes,
in accord with its interpretation as a gauge particle. We also
find that when the bosonic string ends on a D-brane additional negative powers of $\ln M$ multiply the bulk and boundary divergences. These can no longer be cancelled by the ``cosmological constants'',
perhaps pointing to the need, in the presence of D-branes,
for the cancellations of divergences provided by supersymmetry.
\end{abstract}
\vfill
\end{titlepage}
\section{Introduction}
The lightcone parameterization of the string worldsheet
\cite{goddardrt,goddardgrt} provides a framework for
the description of multiloop interacting string diagrams
\cite{mandelstamlc}. The definition of
the lightcone worldsheet path integrals
on a worldsheet lattice \cite{gilest} then provides
a concrete nonperturbative method to study this multiloop expansion
numerically. Monte Carlo methods should be particularly
apt when the diagrams are restricted
to planar open string multiloop diagrams, for which string interactions
decorate the worldsheet lattice in a local manner.
 This restricted
sum of diagrams defines the 't Hooft large $N$ limit
\cite{thooftlargen} of the interacting string theory,
where $N$ is the size of the Chan-Paton matrices
associated with constraining the ends of each open string to move on
a stack of $N$ D-branes. In the case these D-branes
are coincident D3-branes, the
open string spectrum contains a massless $U(N)$ gauge particle
in four spacetime dimensions. This then indicates that
the zero slope limit $\alpha^\prime\to0$ \cite{neveuscherk} of this
sum of diagrams could describe large $N$ QCD \cite{thornsubqcd}.
In this article we restrict our worldsheet lattice
studies to the bosonic string.
We should keep in mind that the bosonic open string tachyon
could make applications to QCD problematic,
either through a failure to stabilize the vacuum or through a stabilization
that breaks the $U(N)$ gauge invariance.
If so, these problems might be
cured by replacing the bosonic open string with
the even G-parity bosonic sector of the
Ramond-Neveu-Schwarz model \cite{ramond,neveuschwarz,gliozziso}.

Given that we will focus on the bosonic string, which has open string
tachyons and hasn't been shown to stabilize, it is necessary, for our studies,
to impose an infrared cutoff that temporarily stabilizes the theory.
As we will describe shortly, there
is a nice way to do this in the context of the worldsheet lattice.
In effect we can naturally impose an energy cost to the existence of each
open string end such that virtual open strings can only exist for
relatively short times. Note that closed string tachyons are not affected
by this infrared cutoff. But closed strings do not propagate
within the planar open string diagrams: in fact their existence is only felt
in their disappearance into the vacuum as described by the holes
in the multiloop worldsheet. Indeed, if we interpret the holes as
closed string emission/absorption by the vacuum, each planar multiloop
diagram can be interpreted
as a closed string tree in a closed string condensate.
Thus the 't Hooft limit just provides us with the subset
of diagrams which might stabilize the vacuum via closed string condensation.
The divergences, which one would normally think of as infrared properties
of the closed string amplitudes, are actually ultraviolet divergences
on the open string worldsheet, which are regulated by the 
worldsheet lattice itself.

Over the last two years, we have been critically
analyzing the continuum limit of the lattice
path integrals for the simplest one open string loop
worldsheets \cite{papathanasiout,papathanasioutwsprop},
and this article is a continuation of these studies.
Our motivation is to clearly understand the
 UV divergence structure emerging from the continuum limit of the lattice
and to determine whether
all UV divergences can be consistently dealt with, either through
cancellation against naturally defined worldsheet counterterms
or through renormalization of the physical
parameters of the theory, the string tension $T_0$ or the 't Hooft
coupling $Ng^2$. Our previous articles
\cite{papathanasiout,papathanasioutwsprop} discussed these issues in the
context of the one open string loop corrections to the closed string
propagator.
Because the only boundary was that of the slit describing the
open string loop, the UV divergences, in this case, arise only from the limit
that the slit length vanishes. For the open string propagator analyzed
in the present article, there are additional UV divergences arising
from the collision of the slit with the boundaries of the open string
worldsheet. The method of reference \cite{papathanasiout},
which took a string field theory approach to the construction of the
one loop propagator proved insufficiently precise to deal
with these boundary divergences. But here we
successfully
apply the worldsheet methods introduced in \cite{papathanasioutwsprop}
to this problem. The key is to represent the lattice
worldsheet propagator in terms of normal modes in discrete time
rather than normal modes in discrete space as was done in
\cite{papathanasioutwsprop} . This makes the
discrete space dependence explicit so that the boundary
contributions to the UV divergence structure can be efficiently
analyzed.

The Giles-Thorn (GT) discretization of the worldsheet \cite{gilest}
begins with a representation of the
free closed or open string propagator as a lightcone worldsheet
path integral defined on a lattice. The lattice replaces the
transverse coordinates of the string ${\bfs x}(\sigma,\tau)$, living on
a rectangular $P^+\times T$ domain, with discretely
labeled coordinates ${\bfs x}_k^j={\bfs x}(kaT_0,ja)$, living
on an $M\times N$ grid with spacing $a$, where $P^+=MaT_0$ and $T=a(N+1)$.
The free string propagator is then simply a Gaussian integral
\bea
{\cal D}_0&=&\int \prod_{kj}d{\bfs x}_k^je^{-S},\nonumber\\
S&=&{T_0\over2}\sum_{kj}\left[({\bfs x}_k^{\ j+1}-{\bfs x}_k^{\ j})^2
+({\bfs x}_{k+1}^{\ j}-{\bfs x}_k^{\ j})^2\right]\equiv
{T_0\over2}{\bfs x}^T\cdot\Delta^{-1}{\bfs x}\,,
\eea
where the $MN\times MN$ matrix
$\Delta$ is the lattice worldsheet propagator.
Then up to an overall
normalization factor ${\cal D}_0={\det}^{-(D-2)}\Delta^{-1}$,
where $D$ is the spacetime dimension ($D=26$ for the bosonic string).

At zero loops, the UV divergences arising in
the continuum limit of the GT
lattice representation of the open and closed
string propagators reside in bulk and boundary
contributions to the ground state energies. The light cone energy
is $P^-=(P^0-P^1)/\sqrt{2}$, and one finds in the continuum
limit \cite{gilest}
\bea
aP^-_{\rm closed,G}&\sim& (D-2)\left[\alpha_0 M-{\pi\over 6M}
+{\mathcal O}(M^{-2})\right]\\
aP^-_{\rm open,G}&\sim& (D-2)\left[\alpha_0 M-\beta_0-{\pi\over 24M}
+{\mathcal O}(M^{-2})\right]
\eea
For the GT lattice one has specifically $\alpha_0=2C/\pi$ and
$\beta_0=\ln(1+\sqrt{2})$ where $C$ is Catalan's constant.
Remembering that $P^+=aMT_0$, we see that
the $1/M$ terms precisely account for the tachyonic masses of the
free closed and open strings. As explained in \cite{gilest}
the $\alpha_0M$ term enters time evolution as an exponential of the combination
$TP^-=(N+1)M\alpha_0$ which is simply proportional to the discretized area
$(N+1)M$ of the lattice: $\alpha_0$ is just a contribution to the
worldsheet bulk ``cosmological constant'' expected in any quantum field
theory. Because the interactions preserve this discrete area,
one can harmlessly introduce a bare bulk cosmological constant $A$ which
is ultimately chosen to cancel all bulk contributions to the
string energies. Similarly the $\beta_0$ can be associated
with the free ends of the open string because it enters the
evolution as an exponential of the combination $-\beta_0(N+1)$
proportional to the length
of the worldsheet boundary. Then we can consistently introduce
a bare worldsheet boundary cosmological constant $B$ chosen to cancel
all these boundary  contributions to the string energies. Unlike
the bulk cosmological constant, this boundary term alters the
dynamics. It is this parameter that provides the infrared cutoff we alluded to
earlier. It is naturally nonzero: even at zero loops
it is necessary to absorb boundary divergences. For the
purposes of our lattice studies we are free to choose it large
enough to suppress the open string tachyonic instability.

On the GT lattice the sum of all
open string multiloop planar diagrams can
be obtained by summing over all patterns of missing spatial
bonds. Formally, this is achieved by introducing Ising-like variables
$S_k^j=0,1$ and taking the worldsheet action to be
\bea\label{action_isinglike}
S_{\rm Planar}&=&{T_0\over2}\sum_{ij}\left[({\bfs x}_i^{\ j+1}
-{\bfs x}_i^{\ j})^2
+S_i^{\ j}({\bfs x}_{i+1}^{\ j}-{\bfs x}_i^{\ j})^2\right]\nonumber\\
&&+(D-2)B\sum_{kj}(1-S_k^j)-\sum_{ij}\left[S_i^{\ j}(1-S_i^{\ j+1})
+S_i^{\ j+1}(1-S_i^{\ j})\right]\ln g\\
&\equiv&{T_0\over2}{\bfs x}^T\cdot\left[\Delta^{-1}+V(S)\right]{\bfs x}
+A(\{S\})\label{lattice_action}\,.
\eea
The terms in $A(\{S\})$ insert the coupling constant $g$ in the
appropriate way and allow for an open string self-energy counterterm
$B$. Then we have
\bea
{\cal D}&=&{\cal D}_0\sum_{\{S\}}{\det}^{-12}(I+V\Delta)e^{-A(\{S\})}\,.
\label{stringfieldprop}
\eea
When $V$ is a sparse matrix, i.e. when there are a relatively small
number of missing bonds (e.g. $\sum_{kj}(1-S_{kj})\ll M$), which
can be arranged by taking $B\gg1$),
this will be a particularly efficient
way to evaluate the terms of perturbation theory. Holding $B$
sufficiently large serves as a physical and convenient infrared
regulator in our studies of the properties of the planar diagrams.
\begin{figure}[ht]
\begin{center}
\includegraphics[width=4in]{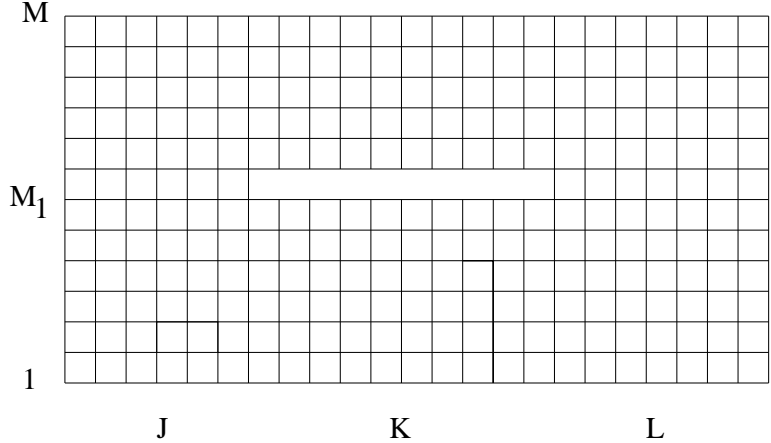}
\caption{\small One loop open string self-energy diagram on the lattice worldsheet. A single open string splits at time $J$ and rejoins at time $J+K$, with total time $N+1=J+K+L\to \infty$ and $J,L \sim N/2$. Thus the diagram is characterized by the number of missing links $K-1$, their position $M_1$, and total string length $M$.}
\label{openselattice}
\end{center}
\end{figure}

The planar open string loop expansion organizes the sum over spins
in (\ref{stringfieldprop}) as a power series in $g^2$, with the number of
loops equal to the number of ``holes'' in the lattice. Orient
the worldsheet so that the time axis ($\tau$) is horizontal and the space
axis ($\sigma$) vertical. Then each hole is
a horizontal row of contiguous missing links. The number of missing
links is the number of time steps the broken string lasts.
In this article we study exclusively one loop corrections
to the open string self-energy,
or, in this language, a single row of contiguous missing links, as we can see in figure \ref{openselattice}.
Since we are concerned here with energy shifts to the free string
spectrum, the initial and final states are energy eigenstates
with the same energy, so we can (and do) take the total number of time steps
$N\to\infty$ keeping the slit's size finite and its
location in the vicinity of $N/2$.
Then a given diagram is characterized by the total
number steps in space (i.e. the number of string bits) $M$,
the length of the slit in lattice units (or number of missing links)
$(K-1)$ and the number of spatial steps $M_1$ between the slit
and one of the open string boundaries. The worldsheet path integral
will depend on $M,K,M_1$, and in principle $K$ should be summed from
1 to $\infty$ and $M_1$ should be summed from 1 to $M-1$. Of
course $M$ is just proportional to the fixed $P^+$ of the
string state whose energy shift is being calculated. The
presence of the open string tachyon renders the $K$ sum exponentially
divergent.

The nature of this tachyonic divergence is easy to see. In
the one loop correction to the open string propagator, the slit
represents the propagation of two open strings as an intermediate
state. The initial and final state is a single open string, say
with (mass)$^2=2\pi(n-1)T_0$ with $n=0,1,2,\ldots$ the mode number of the
state. The intermediate state is two open strings with (mass)$^2=
2\pi(n_1-1)T_0,2\pi(n_2-1)T_0$. If the two open strings last for
a time $(K-1)a$, then the amplitude acquires a factor
$\exp\{-a(K-1)\Delta P^-\}$, where
\bea
a\Delta P^-&=&{\pi(n_1-1)\over M_1}
+{\pi(n_2-1)\over M-M_1}-{\pi(n-1)\over M}\;.
\eea
If $n_1=0$ (or $n_2=0$) $\Delta P^-$ becomes negative for
small enough $M_1$ (or $M-M_1$).  If $M_1$, $M-M_1$, and $M$ are all
of order $M$ in the continuum limit, the coefficient of $(K-1)$ is of
order $M^{-1}$ so as long as $K\ll M$, which is the ultraviolet
region we study here, the exponent stays small. On the other hand
either $M_1$ or $M-M_1$ can be as small as $1$,
in which case the coefficient of $(K-1)$ in the exponential
growth is of order 1, even when $K\ll M$. These large exponential factors
cause practical difficulties with numerical studies, but we will
show that they are absent from the order $M$ and order $M^{-1}$
contributions to the self-energy.

Because the $K$ sum is divergent, we suspend the sum over
$K$ keeping it fixed while we study the large $M$
behavior. The ultraviolet structure that
we wish to analyze is defined by slits much shorter than $P^+$,
or in lattice units $K\ll M$. The continuum limit is $M\to\infty$
so we focus on obtaining the limit of our calculations in the
regime $1\ll K\ll M$. As we contemplate numerical studies
of multiloop diagrams, it is natural to restrict the hole size
summations by simply taking $B$ sufficiently large, rather than
by literally suspending them. With $B$ large enough the tachyonic
instability is stabilized, at the expense of losing Lorentz invariance.
Thus our conclusions strictly apply to this Lorentz
violating cutoff model.

We close this introduction with a brief summary of the results of our
previous work and an outline of the rest of this paper.
In \cite{papathanasiout,papathanasioutwsprop} we analyzed the
one loop correction to the closed string energy. In this case the
sum over $M_1$ is trivial: it just supplies a factor of $M$.
Then the self-energy correction has the form of a single sum over
the slit length $K$
\bea
\Delta P^-&=&\sum_{K=2}^\infty \delta P^-_K
\eea
and we found for $M$ large at fixed $K$
\bea
a\delta P^-_K&\sim& \alpha(K)M+{c(K)\over M}+{d(K)\over M^3}+\cdots
\eea
We determined the large $K$ dependence of the coefficients numerically to
be $\alpha(K)\sim K^{-3}$, and $c(K)\sim K^{-1}$. Thus the coefficient of
$M$ summed over $K$ is finite. This term in the energy is a quadratic
divergence, corresponding to the closed string tachyon. Here we see that
it is in fact harmless\footnote{The harmlessness of
the tachyon divergence in the continuum amplitudes is usually
argued by analytic continuation \cite{GNS}.}. The $K^{-1}$ behavior of the
$1/M$ term signals the UV logarithmic divergence due to the closed
string dilaton. This divergence is, of course, real but can be absorbed in the
slope parameter $\alpha^\prime=1/(2\pi T_0)$. To prove this it is important that
the divergence
is universal for all states. Our work on the closed string
self-energy showed that $c(K)=0$ for the graviton, and had the
appropriate value for selected massive closed string states
to be absorbable into $T_0$.

In this article we deal with the extra complications of the open
string boundaries. In this case the large $M$ expansion at fixed
$K$ has many more terms:
\bea
a\delta P^-_K&\sim& \alpha^\prime(K)M+b(K)+{c^\prime(K)\over M}
+{d^\prime(K)\over M^2}+\cdots
\eea
Here we will show that $\alpha^\prime(K)=\alpha(K)$, necessary
to show the harmlessness of the bulk divergences. We also show that
$c^\prime(K)=c(K)/4$ as required for the consistent absorption of
the logarithmic divergences into the Regge slope parameter. Moreover, we
are able to obtain the large $K$ behavior of the coefficients
analytically using the Fisher-Hartwig formula for the asymptotic
behavior of Toeplitz determinants \cite{FisherHartwig}.
In obtaining these results it is crucial to show that the
exponential divergences, due to an intermediate open string
tachyon with $P^+/(aT_0)={\mathcal O}(1)$, do not contribute to either
the $M$ term or the $1/M$ term. This happens because, before the $M_1$
sum the expansion has the form $a+b/M^2$ and the order $M$ term only arises
by summing $M_1$ over a range of order $M$, and similarly for the
$1/M$ term which comes from summing the $1/M^2$ term over a similar range.
We show that the dangerous exponential divergences contribute only to
the ${\mathcal O}(M^0)$ and higher orders in $1/M$, starting at
${\mathcal O}(M^{-4})$. The constant term can be absorbed in a
renormalization of $B$, but the exponential factors multiplying
the ${\mathcal O}(M^{-4})$ and higher powers of $1/M$
raise practical obstacles to purely numerical efforts to extract the physically
relevant coefficient of $M^{-1}$. Since these obstacles are
directly associated with the open string tachyon instability, there
is at least some hope that if a stabilizing mechanism can be identified,
the numerical difficulties would be surmounted.

In Section 2 we review and generalize the representations for the
worldsheet propagator given in \cite{papathanasioutwsprop}.
We then use these results to analyze the open string self-energy
for the tachyon (Section 3) and the gluon and selected excited states
(Section 4). Then in Section 5 we obtain the large $K$ behavior
of the coefficients in the $1/M$ expansion of the self energies.
Section 6 is devoted to numerical analysis of our results.
Our final Section 7 gives a preliminary discussion of the
problems arising when we try to describe open strings ending on
D-branes, and the possibility that the superstring alleviates them, as indicated by the discretization of the continuum self-energy expressions for the latter.

\section{Worldsheet Propagators}
We gather in this section the expressions for the propagator
on the closed open and Dirichlet worldsheets
found in \cite{papathanasioutwsprop} (see also \cite{discretegreen}).
In that reference the
worldsheet propagator was represented as a spatial normal mode expansion.
But representations based on
temporal normal modes are also useful, so we include them
in our presentation.

Of central interest are the worldsheet correlators of the
coordinates on the $M\times N$
lattice corresponding to the free closed or open string.
\bea
\Delta_{ij,kl}=\VEV{x_i^jx_k^l}={\int {\cal D}x\ x_i^jx_k^l\ e^{-S}\over
\int {\cal D}x\ e^{-S}}
\eea
where the worldsheet action $S$ is appropriate to the type of
string coordinates (closed, open, or Dirichlet) being described.
Because the expectations are taken with Gaussian weight, the
two point correlator in a single dimension captures all of the
relevant information. A straightforward evaluation is to use
closure to write the numerator as the product of three string propagators
(see Appendix \ref{propagators}): one from time $0$ to $j$, one
from times $j$ to $l$, and the last from time $l$ to $+(N+1)$.
We choose Dirichlet boundary conditions in time: $x_i^0=x_i^{N+1}=0$.
We can
resolve $x_i^j$,  $x_k^l$ into spatial normal modes $q_m^j$, $q_n^l$
respectively.
Then because each normal mode integral is independent,
$\VEV{q_m^jq_n^l}=\delta_{mn} \VEV{q_m^jq_m^l}$ one ends up with
a simple two variable Gaussian to do
\bea
\int dq_m^j\ dq_m^l q_m^j q_m^l\exp\left\{-{1\over2}[A_1q_m^{j2}+A_2q_m^{l2})
+2Bq_m^j q_m^l]\right\}&=&-{B\over A_1A_2-B^2}{\det}^{-1/2}
\pmatrix{A_1&B\cr B& A_2\cr}\nonumber\\
\VEV{q_m^jq_n^l}&=&-{B\over A_1A_2-B^2}\delta_{mn}
\eea
Here $A_1,A_2$ and $B$ are read off from the formulas of
Appendix~\ref{propagators}.
We set the $q$'s at the initial and final times to zero.

Then for non-zero modes they are:
\bea
A_1&=&T_0\sinh\lambda\left[\coth j\lambda+\coth(l-j)\lambda\right]\\
A_2&=&T_0\sinh\lambda\left[\coth(N+1-l)\lambda+\coth(l-j)\lambda\right],\qquad
B={-T_0\sinh\lambda\over\sinh(l-j)\lambda}\\
A_1A_2-B^2&=&T_0^2\sinh^2\lambda\left[1+\coth j\lambda\coth(N+1-l)\lambda
\right.\nonumber\\
&&\left.+(\coth j\lambda+\coth(N+1-l)\lambda)\coth(l-j)\lambda\right]\nonumber\\&=&T_0^2\sinh^2\lambda\left[{\sinh(N+1)\lambda\over\sinh j\lambda
\sinh(N+1-l)\lambda\sinh(l-j)\lambda}
\right]\nonumber\\
{-B\over A_1A_2-B^2}&=&{1\over T_0\sinh\lambda}{\sinh j\lambda
\sinh(N+1-l)\lambda\over \sinh(N+1)\lambda}
\eea
where $\lambda$ is $\lambda_m^o$ or $\lambda_m^c$ for the open or closed
string respectively.
For the zero modes
\bea
A_{10}=T_0{l\over j(l-j)},\qquad
A_{20}&=&T_0{(N+1-j)\over(N+1-l)(l-j)},\qquad B_0=-{T_0\over l-j}\nonumber\\
{-B_0\over A_{10}A_{20}-B_0^2}&=&{j(N+1-l)\over T_0(N+1)}
\eea
Then the worldsheet propagator for the open
string worldsheet is given by
\bea
\Delta^o_{ij,kl}&=&{j(N+1-l)\over(N+1)M}+{2\over M}\sum_{m=1}^{M-1}
{1\over\sinh\lambda^o_m}\nonumber\\
&&\times{\sinh j\lambda^o_m
\sinh(N+1-l)\lambda^o_m\over \sinh(N+1)\lambda^o_m}
\cos{m(i-1/2)\pi\over M}\cos{m(k-1/2)\pi\over M},\qquad \hskip-8pt l>j
\label{openpropfsmodes}
\eea
We must keep in mind that this formula applies when $l>j$. In the
opposite
case we switch the roles of $j$ and $l$.
In this formula we have chosen to expand in the normal modes of the
spatial coordinates $i,k$. But we could equally well have chosen
to expand in normal modes in the time coordinates $j,l$. In that case
the propagator takes the form
\bea
\Delta^o_{ij,kl}&=&{2\over N+1}\sum_{n=1}^{N}
{1\over\sinh\lambda^o_n}\nonumber\\&&\times
{\cosh(i-1/2)\lambda^o_n
\cosh(M-k+1/2)\lambda^o_n\over \sinh M\lambda^o_n}
\sin{nj\pi\over N+1}\sin{nl\pi\over N+1},\qquad k>i
\label{openpropftmodes}
\eea
In this case the formula applies when $k>i$. In the
opposite
case we switch the roles of $k$ and $i$.

For string self-energy calculations we want to take $N\to\infty$, but
with $j,l$ well away ($\mathcal{O}(N)$) from $0,N+1$. So to study this
limit we put $j=(N+1)/2+\hat{j}$, $l=(N+1)/2+\hat{l}$, and take the limit
with $\hat{j},\hat{l}$ fixed. Then the two representations take
qualitatively different forms. In the first case we find
\bea
\Delta^o_{ij,kl}&\sim&{N+1\over4M}-{|l-j|\over2M}+{1\over M}\sum_{m=1}^{M-1}
{e^{-|l-j|\lambda^o_m}\over\sinh\lambda^o_m}
\cos{m(i-1/2)\pi\over M}\cos{m(k-1/2)\pi\over M}
\label{openpropsmodes}
\eea
and we see the zero mode divergence in the first term linear in $N$.
In the second case the sum over $n$ turns into an integral and we find
\bea
\Delta^o_{ij,kl}&=&\int_0^1 dx
{\cosh(i-1/2)\lambda^o(x)
\cosh(M-k+1/2)\lambda^o(x)\over\sinh\lambda^o(x)\sinh M\lambda^o(x)}
\cos x({l}-{j})\pi
\label{openproptmodes}
\eea
In this case the zero mode divergence shows up as a divergence in the
integral at the lower limit. In obtaining this formula we used
\be
\sin{nj\pi\over N+1}\sin{nl\pi\over N+1}=\sin^2{n\pi\over2}
\cos{n\hat{j}\pi\over N+1}\cos{n\hat{l}\pi\over N+1}
+\cos^2{n\pi\over2}
\sin{n\hat{j}\pi\over N+1}\sin{n\hat{l}\pi\over N+1}
\ee
The first term contributes only for odd $n$ and the second term only
for even $n$. But in the limit $N\to\infty$ where the sum over $n$ becomes
an integral the right side can be replaced by
\bea
\sin{nj\pi\over N+1}\sin{nl\pi\over N+1}&\to&{1\over2}
\cos{x\hat{j}\pi}\cos{x\hat{l}\pi}
+{1\over2}
\sin{x\hat{j}\pi}\sin{x\hat{l}\pi}
={1\over2}\cos x(\hat{l}-\hat{j})\pi\nonumber\\
&=&{1\over2}\cos x({l}-{j})\pi
\eea
If the open string coordinate satisfies Dirichlet boundary conditions,
the analogs of (\ref{openpropfsmodes}) and (\ref{openpropftmodes})
are
\bea
\Delta^D_{ij,kl}&=&{2\over M}\sum_{m=1}^{M-1}
{1\over\sinh\lambda^o_m}
{\sinh j\lambda^o_m
\sinh(N+1-l)\lambda^o_m\over \sinh(N+1)\lambda^o_m}
\sin{mi\pi\over M}\sin{mk\pi\over M},\qquad \hskip-8pt l>j
\label{Dopenpropfsmodes}
\eea
and
\bea
\Delta^D_{ij,kl}&=&{2\over N+1}\sum_{n=1}^{N}
{1\over\sinh\lambda^o_n}
{\sinh i\lambda^o_n
\sinh(M-k)\lambda^o_n\over \sinh M\lambda^o_n}
\sin{nj\pi\over N+1}\sin{nl\pi\over N+1},\qquad \hskip-8pt k>i
\label{Dopenpropftmodes}
\eea
Correspondingly the analogs of the $N\to\infty$
formulas {} (\ref{openpropsmodes}) and (\ref{openproptmodes}) are
\bea
\Delta^D_{ij,kl}&=&{1\over M}\sum_{m=1}^{M-1}
{e^{-|l-j|\lambda^o_m}\over\sinh\lambda^o_m}
\sin{mi\pi\over M}\sin{mk\pi\over M}
\label{Dopenpropsmodes}
\eea
and
\bea
\Delta^D_{ij,kl}&=&\int_0^1 dx
{\sinh i\lambda^o(x)
\sinh(M-k)\lambda^o(x)\over\sinh\lambda^o(x)\sinh M\lambda^o(x)}
\cos x({l}-{j})\pi
\label{Dopenproptmodes}
\eea
In this case the lower end of the integral shows no divergence, because
zero modes are absent.

For completeness we also mention the two alternative forms for the
worldsheet propagator on the closed string worldsheet.
Expansion in spatial normal modes gives
\bea
\Delta^c_{ij,kl}&=&{j(N+1-l)\over(N+1)M}+{1\over M}\sum_{m=1}^{M-1}
{1\over\sinh\lambda^c_m}\nonumber\\&&\qquad\times
{\sinh j\lambda^c_m
\sinh(N+1-l)\lambda^c_m\over \sinh(N+1)\lambda^c_m}
\exp{2m(i-k){\rm i}\pi\over M},\qquad l>j
\eea
whereas the expansion in temporal normal modes gives
\bea
\Delta^c_{ij,kl}&=&{1\over N+1}\sum_{n=1}^{N}
{1\over\sinh\lambda^o_n}
{\cosh(M/2-|i-k|)\lambda^o_n\over \sinh(M/2)\lambda^o_n}
\sin{nj\pi\over N+1}\sin{nl\pi\over N+1}
\eea
In the first formula we have used Roman ${\rm i}=\sqrt{-1}$
to distinguish it from the index $i$.
Then taking the $N\to\infty$ limit as before leads to, respectively
\bea\label{closed_Delta}
\Delta^c_{ij,kl}
&\sim&{N+1\over 4M}-{|l-j|\over 2M}+{1\over 2M}\sum_{m=1}^{M-1}
{e^{-|l-j|\lambda^c_m}\over\sinh\lambda^c_m}
\exp{2m(i-k){\rm i}\pi\over M}\,.\\
\Delta^c_{ij,kl}&=&{1\over2}\int_0^1 dx
{1\over\sinh\lambda^o_n}
{\cosh(M/2-|i-k|)\lambda^o(x)\over \sinh(M/2)\lambda^o(x)}
\cos x(l-j)\pi\,.
\eea
\section{Open String Tachyon Self-energy}
The one loop self-energy of the ground string
state (the tachyon) can be extracted from the string field
propagator (\ref{stringfieldprop}) by limiting the Ising spin configurations
to those of a single hole of length $K$ (i.e. $K-1$ missing
contiguous missing links), and evaluating the $N\to\infty$ limit
at fixed spin configuration, with the missing links in the vicinity
of time $N/2$. Excited initial and final string states are suppressed
exponentially, so one is left with an amplitude proportional
to the ground string expectation of the interaction, i.e. the
tachyon self-energy times $N$. The proportionality constant is
removed by simply
deleting the factor ${\cal D}_0$ from the expression. The overall
factor of $N$ is removed by fixing the initial time step of the hole
at say $N/2$, so the Ising spin sum is just the sum over
the number of missing links and over the spatial location
of the hole. For the closed string that second sum just provides
a factor of $M$ by spatial translation invariance. But it is
nontrivial for the open string. After all these steps we arrive at the
formula
\bea\label{tachyon_self_energy}
-\Delta P^-&=&g^2\sum_{K=2}^\infty\sum_{M_1=1}^{M-1}
{\det}^{-12}(I+V(M_1,K)\Delta)e^{-24B(K-1)}\,.
\eea
The formula for the closed string tachyon self-energy simplified
because the summand is then independent of $M_1$ so the $M_1$
sum was trivial, leaving only the single sum over $K$.
If $B$ is set equal to its free string value, the sum over $K$ is
badly divergent because the two string intermediate states include
tachyon contributions which are lower in energy than the single string
tachyon. Thus in our studies we are forced to choose $B$ large
enough to regularize this sum. If one could get the answer as
an explicit function of $B$, one could in principle try to continue back to
the free string value. But the perturbation expansion really doesn't make
sense unless the tachyon instability is resolved. What one can do
is study the sum over Ising spins nonperturbatively, holding
$B$ sufficiently large so that the sums over $S$ are convergent, and
then scan the results as a function of $B$ in search of a
meaningful (i.e. Lorentz invariant) result.
\subsection{Self-energy formulas on the continuous worldsheet.}\label{subsection_continuous_selfenergy}
Before doing the worldsheet lattice analysis, we recall the formal continuum
expressions for the open string tachyon and gluon
self-energy in cylinder coordinates, following the notations of \cite{thornsubqcd},
\bea
\Delta P^-_{Tach}&=&{C_o\over 2P^+}\int_0^1{dq\over q^3}{1\over\prod_n(1-q^{2n})^{24}}
\int_0^{2\pi}d\theta{1\over4\sin^2(\theta/2)}
\nonumber\\
&&\hskip2in\prod_{n=1}^\infty
\left({(1-q^{2n}e^{i\theta})(1-q^{2n}e^{-i\theta})
\over(1-q^{2n})^{2}}\right)^{-2}
\label{conttachse}\\
&=&{C_o\over 2P^+}\int_0^1{dq\over q^3}
\int_0^{2\pi}d\theta
\left[{1+24q^2\over4\sin^2(\theta/2)}-2q^2+{\mathcal O}(q^4)\right]
\nonumber\\
\Delta P^-_{Gluon}&=&{C_o\over 2P^+}\int_0^1{dq\over q^3}{1\over\prod_n(1-q^{2n})^{24}}
\int_0^{2\pi}d\theta\left[{1\over4\sin^2(\theta/2)}
-\sum_{n=1}^\infty{2nq^{2n}\over1-q^{2n}}\cos n\theta\right]
\label{contgluonse}\\
&=&{C_o\over 2P^+}\int_0^1{dq\over q^3}
\int_0^{2\pi}d\theta\left[{1+24q^2\over4\sin^2(\theta/2)}
-{2q^{2}}\cos\theta+{\mathcal O}(q^4)\right]\;, \nonumber
\eea
where in each case we displayed the UV behavior $q\sim0$ of the integrand.
The conformal mapping to the lightcone diagram, found in
\cite{mandelstamdet}, determines
the relation of $q,\theta$ to the length $T$ and height $\sigma_1$
of the slit. Interestingly, $\theta$ is simply proportional to
$\sigma_1$, $\theta=2\pi\sigma_1/P^+$ exactly. The relation of $q$
to $T$ is an implicit one involving elliptic functions, which we
give only in the UV limit $q\sim0$:
\bea
q&=&{\pi TT_0\over8P^+\sin\pi\sigma_1/P^+}-{5+\cos2\pi\sigma_1/P^+\over3}
\left({\pi TT_0\over8P^+\sin\pi\sigma_1/P^+}\right)^3+{\mathcal O}(T^5)
\label{qtot}\\
&\to&{\pi K\over8M\sin\pi M_1/M}-{5+\cos2\pi M_1/M\over3}
\left({\pi K\over8M\sin\pi M_1/M}\right)^3+{\mathcal O}(K^5)
\label{qtotdiscrete}\eea
where the second line shows $q$ in the discretized variables
of the lattice, $T=K a$, $\sigma_1=M_1T_0a $.

It is now easy to discretize the self-energy  shift in the UV regime using
\bea
&&{C_o\over 2P^+}\int d\theta\int{dq\over q^3}\to{ C_o\over a\pi T_0M^2}
\sum_{M_1,K}(1+{\mathcal O}(K^4)){64M^2\over K^3}\sin^2{\pi M_1\over M}
\label{qtotjacobian}
\eea
Then for the gluon mass shift we have
\bea
a\Delta P^-_{Gluon}&\to&{16\pi C_o\over T_0}
\sum_{M_1,K}\left({1\over \pi^2K^3}+{3\over8KM^2\sin^2\pi M_1/M}
-{1\over8KM^2}\cos{2\pi M_1\over M}+{\mathcal O}(K^4)\right)\nonumber
\eea
As discussed above, we deal with the severe IR divergences by suspending the
$K$ sum as we study the large $M$ limit:
\bea
a\delta P^-_{Gluon, K}&\to&{16\pi C_o\over T_0}
\Bigg({M-1\over \pi^2K^3}\nonumber\\
&&+{1\over4K}\sum_{M_1=1}^{(M-1)/2}
\left[{3\over M^2\sin^2\pi M_1/M}-{1\over M^2}\cos{2\pi M_1\over M}\right]+{\mathcal O}(K^4)\Bigg)
\eea
The first term is just the familiar bulk term, which we also encountered
for the closed string, and it can be absorbed in the worldsheet cosmological
constant. The first term in square brackets
formally can contribute a physically significant
$1/M$ term, but also an order $M^0$ term\footnote{Contributions to this
constant order $M^0$ term also come from higher terms in the $q$
expansion of the integrand. In general one encounters $M_1$ sums of the
form
\bea
\sum_{M_1=1}^{(M-1)/2}{1\over M^{2n}\sin^{2n}\pi M_1/M}&\sim&
{1\over M^{2n-1}}\int_0^{1/2} dx
\left[{1\over\sin^{2n}\pi x}
-\sum_{k=1}^n{c_k\over x^{2k}}\right]+
\sum_{k=1}^n\sum_{M_1=1}^{(M-1)/2}{c_k\over M^{2(n-k)}M_1^{2k}}\nonumber\\
&\sim& \sum_{k=1}^n{c_k\zeta(2k)\over M^{2(n-k)}}+{\mathcal O}(M^{-2n+1})
={\zeta(2n)\over\pi^{2n}}+{\mathcal O}(M^{-2})\;,\qquad n>1\eea
where the $c_k$ are chosen to make the integral over $x$ finite.
\label{footnote_csc_sums}}. This can be seen as follows:
\bea
\sum_{M_1=1}^{(M-1)/2}{1\over M^2\sin^2\pi M_1/M}&=&
\sum_{M_1=1}^{(M-1)/2}
\left[{1\over M^2\sin^2\pi M_1/M}-{1\over\pi^2M_1^2}\right]+
\sum_{M_1=1}^{(M-1)/2}{1\over\pi^2M_1^2}\nonumber\\
&\sim&{1\over6}-\sum_{m_1=(M+1)/2}^\infty{1\over\pi^2M_1^2}+{1\over M}
\int_0^{1/2} dx\left[{1\over\sin^2\pi x}-{1\over\pi^2x^2}\right]\nonumber\\
&\sim&{1\over6}-{2\over\pi^2(M+1)}+{2\over\pi^2M}
\sim{1\over6}+{\mathcal O}(M^{-2})
\eea
In this case the coefficient of the $1/M$ term is zero! The contribution of the
second term in square brackets involves
\bea
\sum_{M_1=1}^{(M-1)/2}{1\over M^2}\cos{2\pi M_1\over M}&\sim&
{1\over M}\int_0^{1/2} dx \cos2\pi x+{\mathcal O}(M^{-2})={\mathcal O}(M^{-2})
\eea
So in fact there is no $1/M$ contribution to the gluon self-energy,
\bea
a\delta P^-_{Gluon, K}&\to&{16\pi C_o\over T_0}
\left({M-1\over \pi^2K^3}+{1\over8K}+{\mathcal O}(M^{-2})\right)
\eea
consistent with zero mass shift for the gluon.

For the tachyon self-energy the first term in square brackets is the same
as in the gluon self-energy and so gives no contribution to a $1/M$
term. The second term in square brackets, however, when discretized
becomes
\bea
-{C_o a\over P^+}\int {dq\over q}\int d\theta&\to&-{C_o\over MT_0}
\sum_K\sum_{M_1=1}^{M-1}{2\pi\over MK}
\sim-{2\pi C_o\over MT_0}\sum_K {1\over K}
\eea
giving the expected logarithmically divergent tachyon mass shift.
As we shall see in the remainder of this article, the analysis
of the lattice worldsheet is in qualitative accord with
these results.

\subsection{Lattice self-energy, single missing link, $K=2$}
In coordinate space, the matrix $V$ for a single missing link at time $j$ and between spatial positions $k$ and $k+1$
in the open string worldsheet is
\bea\label{Vmatrix}
V_{ml;m^\prime l^\prime}&=&-\delta_{lj}\delta_{l^\prime j}
(\delta_{m,k+1}\delta_{m^\prime,k+1}+\delta_{m,k}\delta_{m^\prime,k}
-\delta_{m,k+1}\delta_{m^\prime k}-\delta_{m^\prime,k+1}\delta_{m k})\,,
\eea
exactly as in the closed string worldsheet. Keeping the propagator
$\Delta$ in coordinate space the necessary determinant of the contributing
$2\times2$ matrix can be taken over from the closed string case:
\bea
\det(I+V\Delta)&=&\det\pmatrix{1+\Delta_{(k+1)j,kj}-\Delta_{kj,kj}
&\Delta_{(k+1)j,(k+1)j}-\Delta_{kj,(k+1)j}\cr
-\Delta_{(k+1)j,kj}+\Delta_{kj,kj}&1-\Delta_{(k+1)j,(k+1)j}+\Delta_{kj,(k+1)j}}
\nonumber\\
&=&1-\Delta_{(k+1)j,(k+1)j}+\Delta_{kj,(k+1)j}+\Delta_{(k+1)j,kj}
-\Delta_{kj,kj}\,.
\label{det1missing}
\eea
We now substitute for $\Delta$ the representation (\ref{openproptmodes})
for the open string worldsheet propagator, which for $l=j$ reduces to
\bea
\Delta^o_{ij,kj}&=&\int_0^1 dx
{\cosh(i-1/2)\lambda^o(x)
\cosh(M-k+1/2)\lambda^o(x)\over\sinh\lambda^o(x)\sinh M\lambda^o(x)},\qquad
k>i
\eea
and we remind the reader that for $k<i$ we switch the roles of $i$ and $k$.
it is helpful to rewrite the numerator in the integrand as
\bea
&&\hskip-1in\cosh(i-1/2)\lambda^o\cosh(M-k+1/2)\lambda^o\nonumber\\
&&={1\over2}[\cosh\lambda^o(M+i-k)+\cosh\lambda^o(M-k-i+1)]\\
&&={1\over2}[\cosh\lambda^oM+\cosh\lambda^o(M-2i+1)],\qquad k=i\\
&&={1\over2}[\cosh\lambda^o(M-1)+\cosh\lambda^o(M-2i)],\qquad k=i+1
\eea
Inserting these results into (\ref{det1missing}), and relabeling $k\to M_1$ to more suitably describe the position of the missing link, leads to
\bea
\det(I+V\Delta^o)&=&\int_0^1dx{\sinh\lambda^o(x)(M-M_1)\sinh\lambda^o(x)M_1
\over\sinh(\lambda^o(x)M/2)\cosh(\lambda^o(x)M/2)}\tanh{\lambda^o(x)\over2}
\eea
Now $\lambda^o(x)=2\sinh^{-1}\sin(\pi x/2)$, and changing integration
variables to $\lambda=\lambda^o$ requires
\bea
{d\lambda\over dx}&=&{\pi\sqrt{1-\sinh^2(\lambda/2)}\over\cosh(\lambda/2)}\;.
\eea
Then we can write
\bea\label{Dk}
{\rm D}_{M_1}\equiv\det(I+V\Delta^o)&=&{1\over\pi}\int_0^{\lambda_0}d\lambda
{\sinh\lambda(M-{M_1})\sinh\lambda {M_1}
\over\sinh(\lambda M/2)\cosh(\lambda M/2)}{\sinh(\lambda/2)\over
\sqrt{1-\sinh^2(\lambda/2)}}
\eea
where $\lambda_0=2\sinh^{-1}1$. It is the value of $\lambda$ where the argument of the square root in the denominator of the integrand vanishes.

We are interested in the limit $M\to\infty$ of the quantity
\bea
-\delta P^-_2&=&\sum_{{M_1}=1}^{M-1} {\rm D}_{M_1}^{-(D-2)}\to\sum_{{M_1}=1}^{M-1} {\rm D}_{M_1}^{-12}=2\sum_{{M_1}<M/2} {\rm D}_{M_1}^{-12}+{\rm D}_{M/2}\delta_{M,even}.
\label{k2tachyon}
\eea
We begin with a study of the large $M$ behavior of ${\rm D}_{M_1}$
itself. The explicit
$M$ dependence of the integrand is buried in the ratio of $\sinh$ and $\cosh$
factors, which for fixed $\lambda>0$ has the behavior
\bea
{\sinh\lambda(M-{M_1})\sinh\lambda {M_1}
\over\sinh(\lambda M/2)\cosh(\lambda M/2)}&\sim&\cases{
1-e^{-2\lambda {M_1}}+{\cal O}(e^{-\lambda M})& for ${M_1}\leq{M\over2}$\cr
\phantom{\int}&\cr
1-e^{-2\lambda(M-{M_1})}+{\cal O}(e^{-\lambda M})& for ${M_1}\geq{M\over2}$\cr}
\eea
Here the exponential terms are included to accurately account
for the cases ${M_1}={\cal O}(1)$, $M-{M_1}={\cal O}(1)$. These terms are as small
as the neglected terms when ${M_1}$ and $M-{M_1}$ are of order $M$.
Next we use $D_{M-{M_1}}=D_{M_1}$ to write the sum over ${M_1}$ in terms of a sum
over ${M_1}\leq M/2$. (If $M$ is odd, it is precisely twice the sum over
${M_1}<M/2$.) Then we break up
\bea\label{integrand_separation}
{\sinh\lambda(M-{M_1})\sinh\lambda {M_1}
\over\sinh(\lambda M/2)\cosh(\lambda M/2)}&=&
1-e^{-2\lambda {M_1}}+\left[-{2e^{-M\lambda}\sinh^2 {M_1}\lambda
\over\sinh M\lambda}\right]
\eea
and evaluate the integral separately for the first two terms and
the term in square brackets:
\bea
{1\over\pi}\int_0^{\lambda_0}d\lambda
(1-e^{-2\lambda {M_1}}){\sinh(\lambda/2)\over
\sqrt{1-\sinh^2(\lambda/2)}}&=&{1\over2}-I_{M_1}\\
I_{M_1}&=&{1\over\pi}\int_0^{\lambda_0}d\lambda
e^{-2\lambda {M_1}}{\sinh(\lambda/2)\over
\sqrt{1-\sinh^2(\lambda/2)}}
\eea
We leave the integral defining $I_{M_1}$ unevaluated, but we will need
its explicit behavior at large ${M_1}$, which can be obtained by expanding
the coefficient of $e^{-2\lambda {M_1}}$ in a power series.
\bea\label{I_{M_1}}
I_{M_1}&=& {1\over\pi}\int_0^\infty d\lambda
e^{-2\lambda {M_1}}\left[{\lambda\over2}+{\lambda^3\over12}+\cdots\right]
+{\cal O}(e^{-2{M_1}\lambda_0})\nonumber\\
&=&{1\over 8\pi {M_1}^2}+ {1\over32\pi {M_1}^4}+{\cal O}({M_1}^{-6})
+{\cal O}(e^{-2{M_1}\lambda_0})
\eea
The exponentially small corrections to this asymptotic
expansion come from the extension of the upper limit
from $\lambda_0$ to $\infty$ used to evaluate the power corrections.

Finally we turn to the contribution of the terms enclosed
in square brackets to ${\rm D}_{M_1}$. By construction it is
exponentially small as $M\to \infty$ at fixed $\lambda$. Thus in
a manner similar to our asymptotic analysis of $I_{M_1}$ we can find
its power behaved large $M$ behavior by expanding its coefficient
in a power series in $\lambda$ and extending the upper limit of integration to
$\infty$. The errors in these steps are exponentially
small:
\bea
{1\over\pi}\int_0^{\lambda_0}d\lambda
\left[\phantom{\bigg|}\right]{\sinh(\lambda/2)\over
\sqrt{1-\sinh^2(\lambda/2)}}&\sim&{1\over\pi}\int_0^{\infty}d\lambda
\left({\lambda\over2M^2}+{\lambda^3\over12M^4}+\cdots\right)
\nonumber\\
&&\times\left[-{2 e^{-\lambda } \sinh ^2(x \lambda )\over \sinh\lambda }\right]\\
&\equiv&{f_2(x)\over2\pi M^2}+{f_4(x)\over12\pi M^4}+\cdots
\eea
where $x\equiv {M_1}/M$. Putting everything together we have
\bea
{\rm D}_{M_1}&=&{1\over2}-I_{M_1}+{f_2(x)\over2\pi M^2}+{f_4(x)\over12\pi M^4}+\cdots
\label{Dklargem}
\eea
We are interested in the large $M$ behavior of
$-\delta P^-\sim aM+b+c/M+\cdots$ through order $1/M$. The sum over ${M_1}$
ranges over $M-1$ values and can thus add up to a power of $M$ to
the explicit $1/M$ dependence of the summand. Thus it is sufficient
to keep only up to order $1/M^2$ in the summand

Inserting these results into (\ref{k2tachyon}) and expanding to the
desired order gives for $M$ odd\footnote{When $M$ is even the
upper limit is $(M-2)/2$ and there is an additional term for ${M_1}=M/2$.}
\bea
-\delta P^-_2&=& 2\sum_{{M_1}=1}^{(M-1)/2}\left[\left({1\over2}-I_{M_1}\right)^{-12}
-{12f_2({M_1}/M)\over M^2}\left({1\over2}-I_{M_1}\right)^{-13}\right]
\eea
Now $I_{M_1}$ is only small at large ${M_1}$ so it is not safe to expand
in powers of $I_{M_1}$. However we can write
\bea
\left({1\over2}-I_{M_1}\right)^{-p}=2^p+2^p\left[\left(1-2I_{M_1}\right)^{-p}-1\right],
\qquad p=12,13
\eea
where the second term behaves as $1/{M_1}^2$ at large ${M_1}$. Because of that
extra convergence the sum over ${M_1}$ does not add a factor of $M$ to the
explicit $1/M$ dependence.
So for the first term in square brackets, the first term is of order
$M$ and the second of order 1:
\bea
2\sum_{{M_1}=1}^{(M-1)/2}\left({1\over2}-I_{M_1}\right)^{-12}&=&
(M-1)2^{12}+2^{13}\sum_{{M_1}=1}^\infty\left[\left(1-2I_{M_1}\right)^{-12}-1\right]
\nonumber\\
&&-2^{13}\sum_{{M_1}=(M+1)/2}^\infty\left[\left(1-2I_{M_1}\right)^{-12}-1\right]\\
&\sim&2^{12}\left[(M-1)+2\sum_{{M_1}=1}^\infty
\left[\left(1-2I_{M_1}\right)^{-12}-1\right]-{12\over\pi M}\right]
\eea
Similarly for the $1/M^2$ term in square brackets, the first term
contributes order $1/M$ but the second term stays of order $1/M^2$:
\bea
-2^{13}{12\over\pi M^2}\sum_{{M_1}=1}^{(M-1)/2}f_2({M_1}/M)\left(1-2I_{M_1}\right)^{-13}
&\sim&-2^{13}{12\over\pi M}\int_0^{1/2}dx f_2(x)
\eea
So we evaluate
\bea
\int_0^{1/2}dx f_2(x)&=&\int_0^{\infty}d\lambda{\lambda}
\int_0^{1/2}dx\left[-{2 e^{-\lambda } \sinh ^2(x \lambda )\over \sinh\lambda }\right]
={\pi^2\over24}-{1\over2}
\eea
So we finally arrive at the large $M$ behavior
\bea\label{deltaP2}
-\delta P^-_2&\sim&2^{12}\left[M-1+2\sum_{{M_1}=1}^\infty
\left[\left(1-2I_{M_1}\right)^{-12}-1\right]-{\pi\over M}\right]+{\cal O}(M^{-2})
\eea
Although for simplicity we assumed that $M$ was odd, it is not difficult
to see that the same result holds for $M$ even.

\subsection{Single slit with $K-1$ missing links}
As we showed in \cite{papathanasioutwsprop}, in the case of a
single slit with $K-1$ missing links between spatial position $k$ and $k+1$,
the path integral (\ref{stringfieldprop}) involves a determinant of the form
\be\label{h_determinant}
\det(I+V\Delta)=\det( h_{lp} )\,,\quad l,p=1,2,\ldots K-1\,,
\ee
where
\be\label{h_elements}
h_{lp}=\delta_{lp}+\Delta_{(k+1)l,kp}-\Delta_{kl,kp}+\Delta_{kl,(k+1)p}-\Delta_{(k+1)l,(k+1)p}\,.
\ee
Focusing on the open string, we insert the two equivalent
 representations (\ref{openpropsmodes}) and (\ref{openproptmodes})
for the propagators,  and again switch to a more distinct notation for the slit position $k\to M_1$, obtaining respectively
\bea
h_{lp}&=&\delta_{lp}-\frac{2}{M}\sum_{m=1}^{M-1}\frac{\sin\frac{m \pi}{2M}\sin^2\frac{m \pi M_1 }{M}}{\sqrt{1+\sin^2 \frac{m \pi}{2M}}}\left(\sin\frac{m \pi}{2M}+\sqrt{1+\sin^2 \frac{m \pi}{2M}}\right)^{-2|l-p|}\label{h_sum}\\
&=&\int_0^{\lambda_0}d\lambda\frac{\sinh \frac{\lambda}{2}\cos\left[2(l-p)\sin^{-1}(\sinh\frac{\lambda}{2})\right]}{\pi\sqrt{1-\sinh^2 \frac{\lambda}{2}}}{\sinh\lambda(M-M_1)\sinh\lambda M_1
\over\sinh(\lambda M/2)\cosh(\lambda M/2)}\,.\label{h_integral}
\eea
In what follows we will  use the integral form
(\ref{h_integral}). Clearly the only difference from (\ref{Dk})
is the additional cosine factor, which carries the dependence on $|l-p|$.
The small $\lambda$ behavior of the first fraction in the
integrand of (\ref{h_integral}) is given by
\be\label{int_common_factor}
\frac{\sinh \frac{\lambda}{2}\cos\left[2(l-p)
\sin^{-1}(\sinh\frac{\lambda}{2})\right]}
{\pi\sqrt{1-\sinh^2 \frac{\lambda}{2}}}
=\frac{\lambda }{2 \pi }
+\frac{\left[1-3 (l-p)^2\right] \lambda ^3}{12 \pi }+\mathcal{O}(\lambda^5)\,.
\ee
Hence we see that contributions to the asymptotic expansion of $h_{lp}$
coming from the ${\cal O}(\lambda)$ term will be the same for one or
many missing links.

Separating the second fraction in the integrand of (\ref{h_integral})
according to (\ref{integrand_separation}), we similarly obtain
\be
h_{lp}=\tilde c_{lp}+\epsilon_{lp}=(c_{lp}-I_{lp})+\epsilon_{lp}\,,
\ee
where
\bea
c_{lp}&=&\int_0^{\lambda_0}d\lambda\frac{\sinh \frac{\lambda}{2}\cos\left[2(l-p)\sin^{-1}(\sinh\frac{\lambda}{2})\right]}{\pi\sqrt{1-\sinh^2 \frac{\lambda}{2}}}=\int_0^{1}dx\frac{\sin \frac{\pi x}{2}\cos\left[(l-p)\pi x\right]}{\sqrt{1+\sin^2 \frac{\pi x}{2}}}\,,\label{c_{lp}}\\
I_{lp}&=&\int_0^{\lambda_0}d\lambda\frac{\sinh \frac{\lambda}{2}\cos\left[2(l-p)\sin^{-1}(\sinh\frac{\lambda}{2})\right]}{\pi\sqrt{1-\sinh^2 \frac{\lambda}{2}}}\,e^{-2M_1\lambda}\,,\label{I_{lp}}\\
\epsilon_{lp}&=&\int_0^{\lambda_0}d\lambda\frac{\sinh \frac{\lambda}{2}\cos\left[2(l-p)\sin^{-1}(\sinh\frac{\lambda}{2})\right]}{\pi\sqrt{1-\sinh^2 \frac{\lambda}{2}}}\frac{2-e^{-2 M_1 \lambda }-e^{2 M_1 \lambda }}{-1+e^{2 M \lambda }}\,.\label{e_{lp}}
\eea
The first part of the matrix element (\ref{c_{lp}})
can be shown to coincide with the $M$-independent part of $h_{lp}$ for the closed string. The $\tilde c_{ij}$ combination,
for $M_1\le (M-1)/2$, encodes the leading behavior of the
integrand for $M$ large.

Expanding as in (\ref{int_common_factor}),
we may formally do the $I_{lp}$ integral term by term using
\be
\int_0^{\lambda_0}\lambda^{s-1} e^{-2M_1\lambda}
=\frac{1}{(2M_1)^s}\int_0^{2M_1\lambda_0}\lambda^{s-1}
e^{-\lambda}=\gamma(s,2M_1\lambda)\,
\ee
where $\gamma(s,x)$ is the lower incomplete gamma function, which
for $s$ a positive integer is
\be
\gamma(s,x)=(s-1)!-(s-1)!e^{-x}\sum_{k=0}^{s-1}\frac{x^k}{k!}\,.
\ee
This expression is particularly useful
for extracting the large $M_1$ behavior of the integral,
since up to exponentially suppressed terms we can write
\be
I_{lp}=\frac{1}{8\pi M_1^2}+\frac{1-3 (l-p)^2}{32 \pi M_1^4}
+\mathcal{O}(M_1^{-6})+\mathcal{O}(e^{-2M_1\lambda})\,.
\ee
In a similar fashion, we obtain a large $M$ expansion for the third integral,
with $x=M_1/M$ arbitrary,
\be
\epsilon_{lp}=\frac{1}{2\pi M^2}f_2(x)
+\frac{1-3 (l-p)^2}{12 \pi M^4}f_4(x)
+\mathcal{O}(M^{-6})+\mathcal{O}(e^{-2M\lambda})\,,
\ee
where $f_i(x)$ are the same functions
that appeared in the single link case\footnote{Neglecting exponentially
suppressed factors, it is not difficult to calculate these
functions explicitly. For example $f_2(x)
=\frac{\pi ^2}{12 M^2}+\frac{1}{4 M^2 x^2}
-\frac{\pi ^2 \csc[\pi  x]^2}{4 M^2}$.}.
Keeping terms of $\mathcal{O}(M^{-2})$ for the
matrix elements will of course yield the determinant
to the same accuracy, and in particular
\be\label{deth_M_expansion}
\det (h_{lp}) =\det \left(\tilde c_{lp}+{\textstyle \frac{f_2(x)}{2\pi M^2}}+\mathcal{O}(M^{-4})\right)=\det (\tilde c_{lp})\left(1+{\textstyle \frac{f_2(x)}{2\pi M^2}}\sum_{l,p=1}^{K-1}(\tilde c^{-1})_{lp}\right)+\mathcal{O}(M^{-4})\,.
\ee
This will be sufficient for obtaining the tachyon self-energy
(\ref{tachyon_self_energy}) for fixed $K$ and $M_1$ summed up to the physically relevant $\mathcal{O}(M^{-1})$ term,
as the latter sum can contribute an extra factor of $M$
at most\footnote{This can be seen, for example,
with the help of the Euler-Maclaurin formula.}.

Finally, another procedure to evaluate the large $M$ expansion of the
sum in question is to add and subtract the value of the
summand for large $M_1$, as we did for the single missing link.
In particular, the analogues of (\ref{I_{lp}})
for the quantities appearing in (\ref{deth_M_expansion}) are
\bea
\det (\tilde c_{lp})&=&\det ( c_{lp})\left(1-{\textstyle \frac{1}{8\pi M_1^2}}\sum_{l,p=1}^{K-1}( c^{-1})_{lp}\right)+\mathcal{O}(M_1^{-4})+\mathcal{O}(e^{-2M_1\lambda})\,,\label{det_ctilde_expansion}\\
\sum_{l,p=1}^{K-1}(\tilde c^{-1})_{lp}&=&\sum_{l,p=1}^{K-1}( c^{-1})_{lp}\left(1+{\textstyle \frac{1}{8\pi M_1^2}}\sum_{l,p=1}^{K-1}( c^{-1})_{lp}\right)+\mathcal{O}(M_1^{-4})+\mathcal{O}(e^{-2M_1\lambda})\,.
\eea
Thus focusing on $M$ odd, we can calculate the self-energy
of a tachyon due to a single slit of $K-1$ time steps
 as follows (in all steps we keep terms up to $\mathcal{O}(M^{-2})$
in the summand or equivalently $\mathcal{O}(M^{-1})$ for the full sum),
\bea
-\delta P^-_K&=& \sum_{M_1=1}^{M-1}\det(h_{lp})^{-12}
=2\sum_{M_1=1}^{(M-1)/2}\det(\tilde c_{lp}+\epsilon_{lp})^{-12}\nonumber\\
&\simeq& 2\sum_{M_1=1}^{(M-1)/2}\det(\tilde c_{lp})^{-12}
\left(1+{\textstyle\frac{f_2(x)}{2\pi M^2}}
\sum_{l,p=1}^{K-1}(\tilde c^{-1})_{lp}\right)^{-12}\nonumber\\
&\simeq& 2\sum_{M_1=1}^{(M-1)/2}\left\{\det( c_{lp})^{-12}+\left(\det(\tilde c_{lp})^{-12}-\det( c_{lp})^{-12}\right)-12{\textstyle\frac{f_2(x)}{2\pi M^2}}\det(\tilde c_{lp})^{-12}\sum_{l,p=1}^{K-1}(\tilde c^{-1})_{lp}\right\}\nonumber\\
&\simeq&(M-1) \det( c_{lp})^{-12}+ 2\sum_{M_1=1}^{\infty}\left(\det(\tilde c_{lp})^{-12}-\det( c_{lp})^{-12}\right)\nonumber\\
&&-2\sum_{M_1=\frac{M+1}{2}}^{\infty}\left(\det(\tilde c_{lp})^{-12}-\det(c_{lp})^{-12}\right)-24\sum_{M_1=1}^{(M-1)/2}{\textstyle\frac{f_2(x)}{2\pi M^2}}\det(\tilde c_{lp})^{-12}\sum_{l,p=1}^{K-1}(\tilde c^{-1})_{lp}\nonumber\\
&\simeq& M \det( c_{lp})^{-12}+ \left[2\sum_{M_1=1}^{\infty}\left(\det(\tilde c_{lp})^{-12}-\det( c_{lp})^{-12}\right)-\det( c_{lp})^{-12}\right]\nonumber\\
&&-24\det( c_{lp})^{-12}\sum_{l,p=1}^{K-1}( c^{-1})_{lp}\left(\sum_{M_1=\frac{M+1}{2}}^{\infty}\frac{1}{8\pi M_1^2}+\sum_{M_1=1}^{(M-1)/2}{\textstyle\frac{f_2(x)}{2\pi M^2}}\right)\nonumber\,.
\eea
For clarity, we mention that in the last step
we dropped the term containing the difference between
$\det( \tilde c_{lp})^{-12}\sum(\tilde c^{-1})_{lp}$
and its asymptotic value in $M_1$,
as it will only contribute at order $\mathcal{O}(M^{-2})$ in the final answer.
Employing the asymptotics
\be
\sum_{M_1=\frac{M+1}{2}}^{\infty}\frac{1}{8\pi M_1^2}=\frac{1}{4\pi M}+\mathcal{O}\left({M^{-2}}\right)\,,\quad \sum_{M_1=1}^{(M-1)/2}\frac{f_2(x)}{2\pi M^2}=-\frac{1}{4 M \pi }+\frac{\pi }{48 M}+\mathcal{O}\left(M^{-3}\right)\,,
\ee
we finally obtain
\bea\label{deltaP2_single_slit}
-\delta P^-_K&=& M \det( c_{lp})^{-12}+ \left[2\sum_{M_1=1}^{\infty}\left(\det(\tilde c_{lp})^{-12}-\det( c_{lp})^{-12}\right)-\det( c_{lp})^{-12}\right]\nonumber\\
&&\,\,\,-\frac{\pi}{2M}\det( c_{lp})^{-12}\sum_{l,p=1}^{K-1}( c^{-1})_{lp}+\mathcal{O}\left({M^{-2}}\right)\,.
\eea
Comparing with the respective summand for the closed string,
see equations (38) and (59) in \cite{papathanasioutwsprop},
we note that the leading term in the two expressions is the same,
and the $\mathcal{O}(M^{-1})$ is four times larger in the closed string.
The same proportionality holds between the tachyon masses of the
free closed and open strings.

\section{Open String Gluon Self-energy}
To extract energy shifts for excited states we examine the
propagator on a lattice worldsheet with some pattern of missing links
described by $V$:
\bea\label{propagator_interacting}
\Delta^V&=&(\Delta^{-1}+V)^{-1}=\Delta(I+V\Delta)^{-1}
=\Delta-\Delta(I+V\Delta)^{-1}V\Delta\equiv\Delta-\Delta{\cal V}\Delta\,.
\eea
Inserting the normal mode expansion (\ref{openpropsmodes}) for $\Delta$
in the rightmost side of this equation we can write $\Delta^V$:
\bea
\Delta^V_{ij,kl}&=&\sum_{m,m^\prime}e^{j\lambda^o_m-l\lambda^o_{m^\prime}}
{{\tilde\Delta}^V_{mm^\prime}\over M
\sqrt{\sinh\lambda^o_m\sinh\lambda^o_{m^\prime}}}
\cos{m(i-1/2)\pi\over M}\cos{m^\prime(k-1/2)\pi\over M}\\
{\tilde\Delta}^V_{mm^\prime}&=&\delta_{mm^\prime}-{{\tilde{\cal V}}_{mm^\prime}
\over M\sqrt{\sinh\lambda^o_m\sinh\lambda^o_{m^\prime}}}\\
{\tilde{\cal V}}_{mm^\prime}&=&\sum_{pq,rs}{{\cal V}}_{pq,rs}
e^{-q\lambda^o_m+s\lambda^o_{m^\prime}}
\cos{m(p-1/2)\pi\over M}
\cos{m^\prime(r-1/2)\pi\over M}\label{calV_tilde}
\eea
then the contribution of this diagram to the one loop gluon self-energy is
\bea
-{\tilde\Delta}^V_{11}{\det}^{-12}(I+V\Delta)
\eea
where $V$ corresponds to the missing link patterns of a single
hole in the worldsheet.
\subsection{Single missing link, $K=2$}
Working in the $2\times2$ subspace selected by $V$ for a single missing link between positions $k, k+1$ at time $j$,
we have, putting $A=\Delta_{(k+1)j,kj}-\Delta_{kj,kj}$ and
$A^\prime=\Delta_{(k+1)j,kj}-\Delta_{(k+1)j,(k+1)j}$,
\bea
V&=&\pmatrix{-1&1\cr1&-1\cr},\qquad
I+V\Delta=\pmatrix{1+A&-A\cr -A^\prime&1+A^\prime\cr}\,,\nonumber\\
 {\cal V}=(I+V\Delta)^{-1}V&=&
{1\over1+A+A^\prime}\pmatrix{1+A^\prime&A\cr A^\prime&1+A\cr}
\pmatrix{-1&1\cr1&-1\cr}={V\over1+A+A^\prime}\nonumber\\
&=&V{\det}^{-1}(1+V\Delta)\,.
\eea
Then
\bea
{\tilde{\cal V}}_{mm^\prime}&=&
\bigg[\cos{m(k-1/2)\pi\over M}
\cos{m^\prime(k+1/2)\pi\over M}+\cos{m(k+1/2)\pi\over M}
\cos{m^\prime(k-1/2)\pi\over M}\nonumber\\
&&\hskip-0.8in-\cos{m(k-1/2)\pi\over M}
\cos{m^\prime(k-1/2)\pi\over M}-\cos{m(k+1/2)\pi\over M}
\cos{m^\prime(k+1/2)\pi\over M}\bigg]{e^{-j(\lambda^o_m-\lambda^o_{m^\prime})}\over\det(1+V\Delta)}\nonumber\\
&=&-4\sin{m\pi\over 2M}\sin{m^\prime\pi\over 2M}\sin{mk\pi\over M}
\sin{m^\prime k\pi\over M}{e^{-j(\lambda^o_m-\lambda^o_{m^\prime})}
\over\det(1+V\Delta)}\\
{\tilde\Delta}^V_{mm^\prime}&=&\delta_{mm^\prime}+4
{\sin{m\pi\over 2M}\sin{m^\prime\pi\over 2M}\sin{mk\pi\over M}
\sin{m^\prime k\pi\over M}
\over M\sqrt{\sinh\lambda^o_m\sinh\lambda^o_{m^\prime}}}
{e^{-j(\lambda^o_m-\lambda^o_{m^\prime})}
\over\det(1+V\Delta)}
\eea
Then the contribution to the self-energy of the excited string state
$a_{-m}\ket{0}$ from this diagram (setting $m^\prime=m$)
is
\bea
-\delta P^-_2(m)&=&\sum_{k=1}^{M-1}\left(1+4
{\sin^2(\frac{m\pi}{2M})\sin^2(\frac{mk\pi}{M})
\over M\sinh\lambda^o_m}
{1
\over\det(1+V\Delta)}\right){\det}^{-12}(1+V\Delta)\\
&\approx&\sum_{k=1}^{M-1}\left(1+
{m\pi\over M^2}
{\sin^2(\frac{mk\pi}{M})
\over\det(1+V\Delta)}+{\cal O}(M^{-4})\right){\det}^{-12}(1+V\Delta)
\nonumber\\
&\sim&2^{12}\left[M-1+2\sum_{k=1}^\infty
\left[\left(1-2I_k\right)^{-12}-1\right]+(m-1){\pi\over M}\right]+{\cal O}(M^{-2})\label{deltaP2_gluon}
\eea
It is of course significant that the coefficient $1/M$ vanishes for $m=1$,
reflecting the fact that perturbative corrections to the
gluon mass should be 0.

\subsection{Single slit with $K-1$ missing links}
As we've shown in \cite{papathanasioutwsprop}, and can directly verify from the definition
\be
\mathcal{V}\equiv(I+V\Delta)^{-1}V\,\Rightarrow (I+V\Delta)\mathcal{V}=V\,,
\ee
the elements of $\mathcal{V}$ are given by
\be\label{calV_elements}
\mathcal{V}_{kl,ks}=\mathcal{V}_{(k+1)l,(k+1)s}=-\mathcal{V}_{(k+1)l,ks}=-\mathcal{V}_{kl,(k+1)s}=-h^{-1}_{ls}\,,
\ee
where $k$ denotes the spatial position of the slit, and the matrix $h$ was defined in (\ref{h_sum})-(\ref{h_integral}). In what follows, we will again redefine $k\to M_1$ so as to label the slit position in a more distinctive manner. In this notation, and with the help of (\ref{calV_elements}), the Fourier transform of $\mathcal{V}$ with the open string wavefunctions (\ref{calV_tilde}) becomes
\be
\tilde \mathcal{V}_{m m^\prime}=-4\sin{m\pi\over 2M}\sin{m^\prime\pi\over 2M}\sin{m{M_1}\pi\over M}
\sin{m^\prime {M_1}\pi\over M}\sum_{q,s=1}^{K-1}e^{-s \lambda^o_m+q \lambda^o_{m^\prime}} h^{-1}_{qs}\,.
\ee
Then the analogue of (\ref{deltaP2_gluon}) for many missing links will be
\bea\label{deltaP_gluon}
-\delta P^-_{K}(m)&=&\sum_{{M_1}=1}^{M-1}{\tilde\Delta}^V_{mm}{\det}^{-12}(I+V\Delta)=\sum_{{M_1}=1}^{M-1}\left(1-\frac{\tilde\mathcal{V}_{mm}}{M\sinh\lambda^o_m}\right){\det}^{-12}(h_{lp})\\
&\approx&\sum_{{M_1}=1}^{M-1}\left(1+
{m\pi\over M^2}
{\sin^2(m\pi{M_1}/M)}\sum_{q,s}e^{(q-s) \lambda^o_{m}} h^{-1}_{qs}+{\cal O}(M^{-4})\right){\det}^{-12}(h_{lp})\,,
\nonumber
\eea
The additional contribution for the gluon as compared to the tachyon comes from the second term in the parenthesis in (\ref{deltaP_gluon}). We are interested in the asymptotic expansion of the latter equation only up to $\mathcal{O}(M^{-1})$, and hence we only need the leading term of the additional contribution. As we discussed in the case of the tachyon, this may be obtained by replacing all quantities in the sum in ${M_1}$ with their asymptotic form for large ${M_1}$, which for the case at hand implies
\be
\left(\sum_{q,s}e^{(s-q) \lambda^o_{m}} h^{-1}_{qs}\right){\det}^{-12}(h_{lp})\rightarrow \left(\sum_{q,s} c^{-1}_{qs}\right){\det}^{-12}(c_{lp})
\ee
Then, the sum in ${M_1}$ can be done exactly in terms of geometric series, and together with the contribution which is identical to the tachyon self-energy (\ref{deltaP2_single_slit}), we obtain the final formula
\bea\label{deltaP_single_slit}
-\delta P^-_K(m)&=& M \det( c_{lp})^{-12}+ \left[2\sum_{M_1=1}^{\infty}\left(\det(\tilde c_{lp})^{-12}-\det( c_{lp})^{-12}\right)-\det( c_{lp})^{-12}\right]\nonumber\\
&&\,\,\,+(m-1)\frac{\pi}{2M}\det( c_{lp})^{-12}\sum_{l,p=1}^{K-1}( c^{-1})_{lp}+\mathcal{O}\left({M^{-2}}\right)\,.
\eea

\section{Dependence of String Self-energy on Slit Size}\label{sec_slit_size_dependence}

\subsection{Leading term in the $M$ expansion via Fisher-Hartwig formula}
We would like to know the dependence of the coefficients
of (\ref{deltaP_single_slit}) on the slit size $K-1$,
for $M \gg K\gg1$. At first this seems quite challenging,
as the dependence on $n=K-1$ enters primarily via the size
of the $\det (c_{lp})$ determinant.

Fortunately, this can be achieved by exploiting the fact
that the latter is the determinant of a Toeplitz matrix,
meaning that $c_{lp}=c(l-p)$, or in other words
that all elements in a left-to-right descending diagonal are the same.
In this case, there exists a formula for
the asymptotic behavior of the determinant
due to Fisher and Hartwig \cite{FisherHartwig},
see also \cite{2012arXiv1206.1292D} for a more recent treatment.
We will rely on the notations of the latter paper,
and in particular we will rewrite the matrix elements as
Fourier transforms of the same function $f(z)$,
\be\label{toeplitz_element}
c_{lp}=\int_0^{1}dx\frac{\sin({\pi x}/{2})
\cos\left[(l-p)\pi x\right]}{\sqrt{1+\sin^2({\pi x}/{2})}}=\frac{1}{2\pi}\int_0^{2\pi}d\theta f(e^{i\theta})e^{-i(l-p)\theta}\,,\quad l,p=1,\ldots,n
\ee
where
\be
f(e^{i\theta})=\frac{\sin({\theta}/{2})}{\sqrt{1+\sin^2({\theta}/{2})}}\,
\ee
or equivalently, for $z=e^{i \theta}$
\be
f(z)=\frac{|z-1|}{2\sqrt{1+(z-1)(1/z-1)/4}}\,.
\ee
This implies that $f(z)$ is a special case of the function
considered in \cite{2012arXiv1206.1292D}, with the following values
for the parameters according to their conventions,
\be
z_0=1\,,\quad\alpha_0=\frac{1}{2}\,,
\quad\beta_0=0\,,\quad V(e^{i\theta})
=-\log\left(2\sqrt{1+{\textstyle \sin^2 \frac{\theta}{2}}}\right).
\ee
Consequently, the asymptotic behavior of the $n$-dimensional determinant will be given by
\be\label{detc_asymptotic_initial}
\det (c_{lp})=n^{\frac{1}{4}}\exp\left(n V_0+\sum_{k=1}^\infty k V_k V_{-k}-\frac{1}{2}\sum_{k=1}^\infty V_k-\frac{1}{2}\sum_{k=-\infty}^{-1} V_{k}\right)\frac{G(\frac{3}{2})^2}{G(2)}\left(1+\mathcal{O}(n^{-1})\right)\,,
\ee
where $G(x)$ is the Barnes G-function and $V_k$ are the Fourier modes of $V(e^{i\theta})$,
\be
V_k=\frac{1}{2\pi}\int_0^{2\pi}d\theta V(e^{i\theta})e^{-ik\theta}\,,\quad V(z)=\sum_{k=-\infty}^{\infty}V_k z^k\,.
\ee
Our function $V(e^{i\theta})$ is simple enough that $V_{-k}=V_k$, and in particular we can calculate them exactly,
\be
V_0=-\log(1+\sqrt{2})\,,\quad V_k=\frac{1}{2k}(1-\sqrt{2})^{2k}\,
\ee
from which we can in turn obtain
\be
\sum_{k=1}^\infty k V_k V_{-k}=-\frac{1}{4} \log\left[4 (-4+3 \sqrt{2})\right]\,,\quad\sum_{k=1}^\infty V_k=\sum_{k=1}^\infty V_{-k}=-\frac{1}{2} \log \left[2 (\sqrt{2}-1)\right]\,.
\ee
Substituting back into (\ref{detc_asymptotic_initial}), we obtain the final formula
\bea\label{detc_asymptotic_final}
\det (c_{lp})&=&n^{\frac{1}{4}}\exp\left(-\log(1+\sqrt{2})n -\frac{1}{8} \log 2\right)\frac{G(\frac{3}{2})^2}{G(2)}\left(1+\mathcal{O}(n^{-1})\right)\,,\nonumber\\
&\simeq&\exp\left(0.25\log n-0.881n+0.0472\right)\left(1+\mathcal{O}(n^{-1})\right)\,,\\
&\simeq&1.048 \,n^{
\frac{1}{4}}\exp\left(-0.881n\right)\left(1+\mathcal{O}(n^{-1})\right)\,.\nonumber
\eea
As a consistency check, we can compare the asymptotic formula above with fits for the value of the determinant over a range of different $n$. For this purpose, it turns more efficient to fit the logarithm of the determinant, and we choose the range $n\in [100,200]$ in steps on 1. We find that
\be
\log \det (c_{lp})\simeq 0.2499 \log n - 0.88137 n +0.0472 + \frac{0.17}{n}
\ee
where the errors in the coefficients are at the order of the last digit, and we also included a term $\log(1+c/n)\simeq c/n$ to account for the subleading asymptotic term in (\ref{detc_asymptotic_final}). Evidently the coefficients of the fit are in excellent agreement with the Fisher-Hartwig formula.

In order to obtain the dependence of the leading term in the $M$-expansion of the tachyon and gluon self-energy summand\footnote{In fact, this term is universal for all states of the open and closed string.} on the duration of the self-interaction $n=K-1$, we simply have to raise (\ref{detc_asymptotic_final}) to the $(-12)$ power. In this manner we obtain a power dependence of $n^{-3}$, which is a rigorous confirmation of the rough estimate we had obtained in \cite{papathanasiout}.

\subsection{$\mathcal{O}(M^{-1})$ term}
In order to find the dependence on slit size for the $\mathcal{O}(M^{-1})$ term in (\ref{deltaP_single_slit}), we will have to additionally analyze $\sum c^{-1}_{lp}$.  To this end, we will be using asymptotic expansions for the inverses of Toeplitz matrices in the same category with $c_{lp}$, which have relatively recently appeared in the literature \cite{Rambour2009489,Rambour2010155}.

In more detail, these papers focus on Toeplitz matrices of the form
(\ref{toeplitz_element}), where
\be
f(z)=|z-1|^{2\alpha} f_1(z)\,,
\ee
so that our matrix of interest, $c_{lp}$, is a special case with $\alpha=1/2$ and
\be
f_1(z)=\frac{1}{2\sqrt{1+(z-1)(1/z-1)/4}}\,.
\ee
In order to stay as close as possible to the notations of \cite{Rambour2009489,Rambour2010155}, let us call the dimensionality of the matrix $n\equiv N+1$. Then, for $0<x<1$, $0<y<1$, $x\ne y$, the asymptotic forms of the inverse matrix element will be
\bea
c^{-1}_{[Nx]+1,1}&=&\frac{1}{g_1(1)\sqrt{\pi N} }\frac{\sqrt{1-x}}{\sqrt{x}}+o(N^{-1/2})\,,\qquad f_1=g_1\bar g_1\,,\label{c_inverse_asymptotic_boundary}\\
c^{-1}_{[Nx]+1,[Nx]+1}&=&\frac{1}{f_1(1)\pi }\log N+o(\log N)\,,\label{c_inverse_asymptotic_diagonal}\\
c^{-1}_{[Nx]+1,[Ny]+1}&=&\frac{1}{f_1(1)\pi}G_{\frac{1}{2}}(x,y)+o(1)\,,\label{c_inverse_asymptotic}
\eea
where $[a]$ denotes the integer part of $a$,
\be
G_{\frac{1}{2}}(x,y)=\sqrt{x}\sqrt{y}\int^1_{\max(x,y)}\frac{dt}{t\sqrt{t-x}\sqrt{t-y}}=2\textrm{arctanh}\frac{\sqrt{y}}{\sqrt{1-y}}\frac{\sqrt{1-x}}{\sqrt{x}}\,,
\ee
and the last equality in the above equation holds if $x>y$, otherwise we simply exchange $x\leftrightarrow y$.

\begin{figure}
\begin{center}
\includegraphics[width=5in]{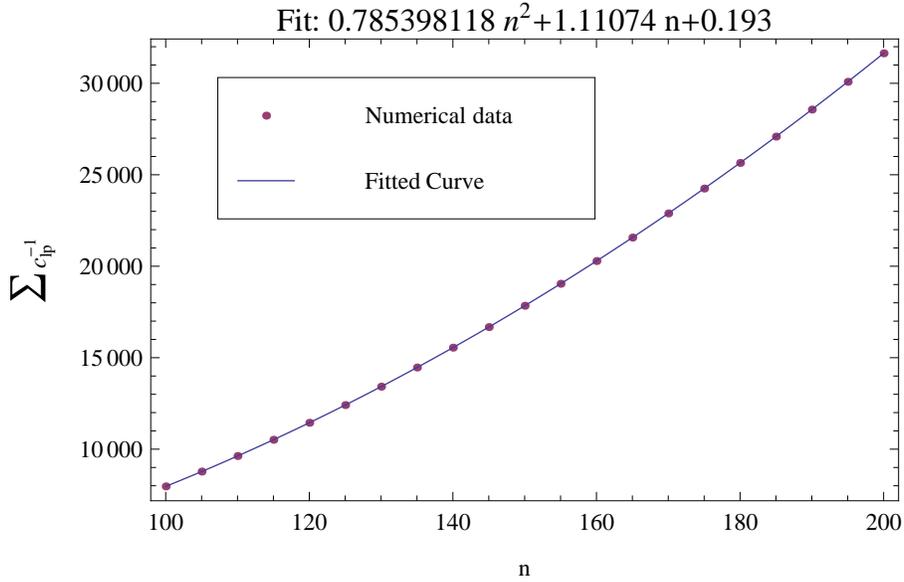}
\caption{Polynomial fit of degree two for the quantity $\sum_{l,p}^n(c^{-1})_{lp}$, appearing in the $\mathcal{O}(M^{-1})$ term in the asymptotic expansion for one loop tachyon self-energy (\ref{deltaP2_single_slit}). The coefficient of the leading $\mathcal{O}(n^2)$ term agrees excellently with the analytically derived value $\pi/4$.\label{cfit}}
\end{center}
\end{figure}

According to the Euler-Maclaurin formula, the leading contribution to the sum over the $[Nx]$ or $[Ny]$ indices in each of the formulas above will be $N$ times the integral over $x$ or $y$. This implies a contribution of order $\mathcal{O}(N^{1/2})$, $\mathcal{O}(N\log N)$ and $\mathcal{O}(N^2)$ to the sum over all indices from (\ref{c_inverse_asymptotic_boundary}), (\ref{c_inverse_asymptotic_diagonal} and (\ref{c_inverse_asymptotic}) respectively. So up to leading order we may write
\bea
\sum_{l,p=1}^{N+1}( c^{-1})_{lp}&\simeq&\sum_{l\ne p=1}^{N+1}( c^{-1})_{lp}=2\sum_{l> p=1}^{N+1}( c^{-1})_{lp}\nonumber\\
&\simeq&\frac{2N^2}{f_1(1)\pi}\int_0^1 dx\int_0^x dy G_{\frac{1}{2}}(x,y)\\
&\simeq&\frac{4N^2}{\pi}\int_0^1 dx\left[2 \sqrt{x(1-x)}\arcsin\sqrt{y}+2 (y-x) \textrm{arctanh}\frac{\sqrt{1-x} \sqrt{y}}{\sqrt{x} \sqrt{1-y}}\right]_{y=0}^x\nonumber\\
&\simeq&\frac{8N^2}{\pi}\int_0^1 dx \sqrt{x(1-x)} \arcsin\sqrt{x}\nonumber\,,
\eea
and given that the $x$-integral above yields $\pi^2/32$, we finally obtain, after we restore $n=N+1$,
\be\label{c_inverse_sum}
\sum_{l,p=1}^{n}( c^{-1})_{lp}\simeq \frac{\pi}{4}\,n^2\simeq 0.78539816\ n^2\,.
\ee
Another way to arrive at this result,
is to notice that only the value of $f_1(z)$ at $z=1$ matters
for the leading term in the expansions
(\ref{c_inverse_asymptotic_boundary})-(\ref{c_inverse_asymptotic}). In order to extract the term in question, we can thus examine the determinant where we have replaced $f_1(z)$ with its constant value $f_1(1)$, namely
\be
d_{lp}=\frac{1}{2\pi}\int_0^{2\pi}d\theta \sin \frac{\theta}{2}e^{-i(l-p)\theta}=\frac{2}{\pi}\frac{1}{1-4(l-p)^2}\,.
\ee
Evidently, the virtue of this replacement is that it allows us to compute the integral explicitly. Then, by analytically inverting the matrix and summing its elements for $n\in[1,10]$, we experimentally find that the sum of all elements of the inverse matrix is given by the following simple formula,
\be
\sum_{l,p=1}^{n}( d^{-1})_{lp}= \frac{\pi}{4}\,n(n+1)\,.
\ee
As a final test of our result (\ref{c_inverse_sum}), we may compare it to fits of the quantity for varying $n$. In particular, we choose $n\in[100,200]$ in steps of 5, and determine the coefficients of a polynomial fit of degree two, as potentially existing logarithms at orders lower than $\mathcal{O}(n^2)$ can be well approximated by constants within this range. The results are depicted in Figure \ref{cfit}, and leave no doubt that the leading dependence of the sum of all elements of matrix $c^{-1}$ on its size is given by (\ref{c_inverse_sum}).
\section{Numerical Analysis}
\begin{table}
\begin{tabular}{|c|c|c|}\hline
$K$&$-\delta P^-_{\rm{Tachyon}} \textrm{ fit}$&$-\delta P^-_{\rm{Tachyon}} \textrm{ asymptotic formula}$\\\hline
 2 & $\scriptstyle 4096 M+19804-12867.90/M $& $\scriptstyle4096 M+19803-12867.96/M $\\
 3 & $\scriptstyle 4.2569937517\times 10^7 M+1.48720\times 10^9-3.68032\times 10^8/M$ & $\scriptstyle4.2569937516\times 10^7 M+1.48717\times 10^9-3.68037\times 10^8/M$ \\
 4 & $\scriptstyle6.602641227\times 10^{11} M+1.63308\times 10^{14}-1.09497\times 10^{13}/M$ & $\scriptstyle6.602641228\times 10^{11} M+1.63307\times 10^{14}-1.09501\times 10^{13}/M $\\
 5 & $\scriptstyle1.2725545528\times 10^{16} M+2.743318\times 10^{19}-3.4330\times 10^{17}/M$ & $\scriptstyle1.2725545522\times 10^{16} M+2.743315\times 10^{19}-3.4332\times 10^{17}/M$\\\hline
\end{tabular}
\caption{Tachyon self-energy for an interaction lasting $K-1$ time steps, $K=2,\ldots,5$. Coefficients of asymptotic expansion in $M$ up to $\mathcal{O}(M^{-1})$, as obtained by numerically evaluating and fitting the quantity in question for $M\in[1005,1995]$ in steps of 10, and compared to formula (\ref{deltaP2_single_slit}).\label{table_deltaP_comparison}}
\end{table}
In this section, we numerically evaluate the tachyon and gluon self-energy
$\delta P^-_K$ due to an interaction lasting $K-1$ time steps,
investigate its behavior for different values of $M$ and $K$,
and obtain fits that we compare to the analytic asymptotic formulas
(\ref{deltaP2_single_slit}), (\ref{deltaP_single_slit}). We are interested in the ultraviolet $M\gg K$
behavior of the self-energy, so we choose $K\in{[2,30]}$ and
$M\in{[495,1995]}$
in steps of 10. We perform the evaluation using
the sum expression for the matrix elements of the determinant (\ref{h_sum}),
as it turns out to be numerically
more stable than the integral expression (\ref{h_integral}).

\begin{figure}
\centering
\begin{subfigure}[b]{0.49\textwidth}
\includegraphics[width=\textwidth]{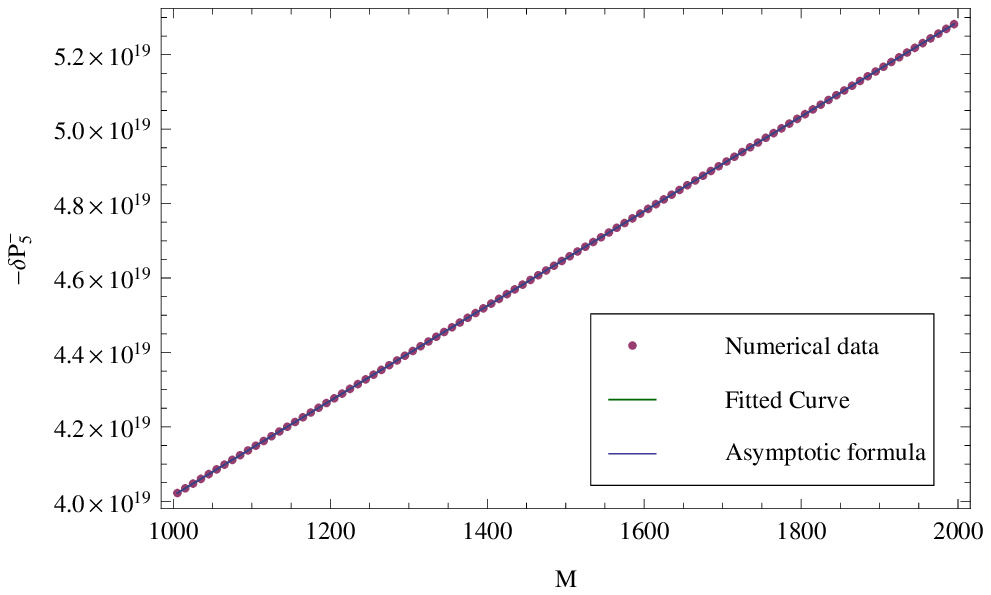}
\caption{\label{plot_tachyon_k5}}
\end{subfigure}
\begin{subfigure}[b]{0.49 \textwidth}
\includegraphics[width=\textwidth]{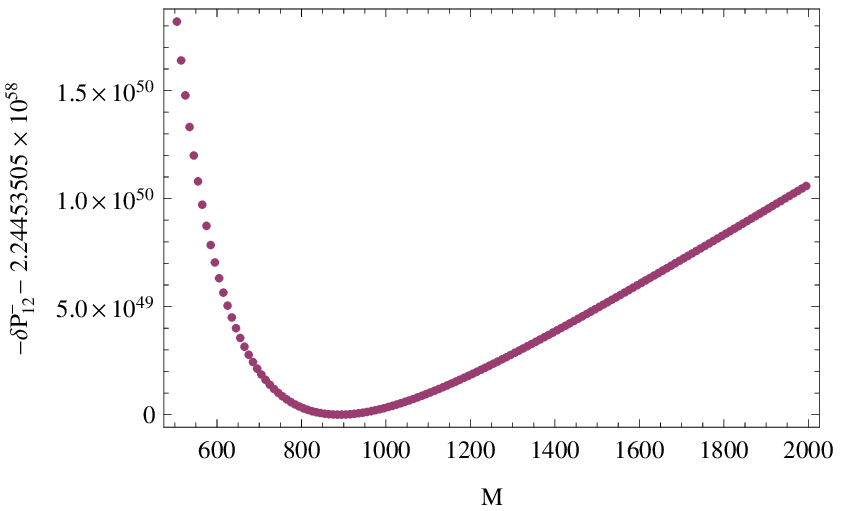}
\caption{\label{plot_tachyon_k12}}
\end{subfigure}
\caption{Plots of tachyon self-energy for an interaction lasting $K-1$ time steps, for $K=5$ in (a) and $K=12$ in (b). Whereas for (a) the $\mathcal{O}(M)$ term dominates and the data, fit and asymptotic formula up to $\mathcal{O}(M^{-1})$ are indistinguishable, the same does not hold for the lower end of $M$ values in (b).}
\label{fig:DeltaP_K}
\end{figure}

Starting with the tachyon, we observe that for the first few values of fixed $K$ the leading behavior
of $\delta P_K^-$ is indeed linear in $M$ within the range we have chosen,
and performing fits of the form $\delta P^-_K=\sum c_i M^i$ with three
free parameters $c_{\pm1}, c_{0}$, we find excellent agreement of the
values predicted by (\ref{deltaP2_single_slit}). The results of the
fits and the comparison with the asymptotic prediction are depicted
in Table \ref{table_deltaP_comparison}, see also Figure \ref{plot_tachyon_k5}.

As it is evident in Figure \ref{plot_tachyon_k12} however, starting at $K=12$ and higher $-\delta P^-_K$ becomes convex, and subleading terms in the $M$ expansion begin to dominate over the linearly increasing term in the lower end of our range. In particular, this behavior at lower $M$ cannot be due to the $\mathcal{O}(M^{-1})$ term, which could only cause a deviation below the straight line because of its negative sign. Therefore it must come from higher terms in the expansion, and experimentation with different fitting functions suggests that it is in fact due the $\mathcal{O}(M^{-4})$ term.
\begin{figure}
\centering
\begin{subfigure}[b]{0.49\textwidth}
\includegraphics[width=\textwidth]{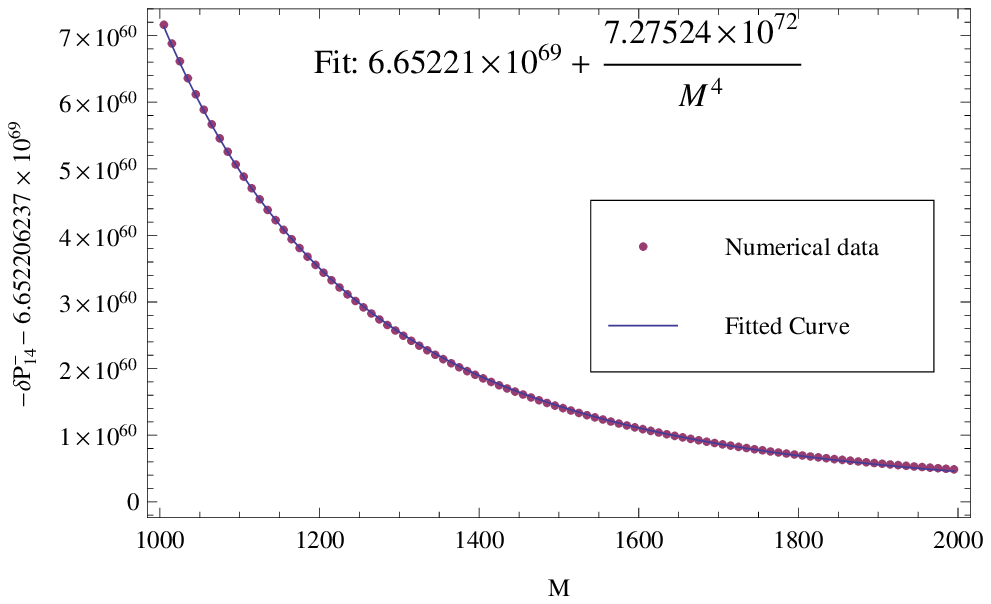}
\caption{\label{plot_tachyon_k14}}
\end{subfigure}
\begin{subfigure}[b]{0.49 \textwidth}
\includegraphics[width=\textwidth]{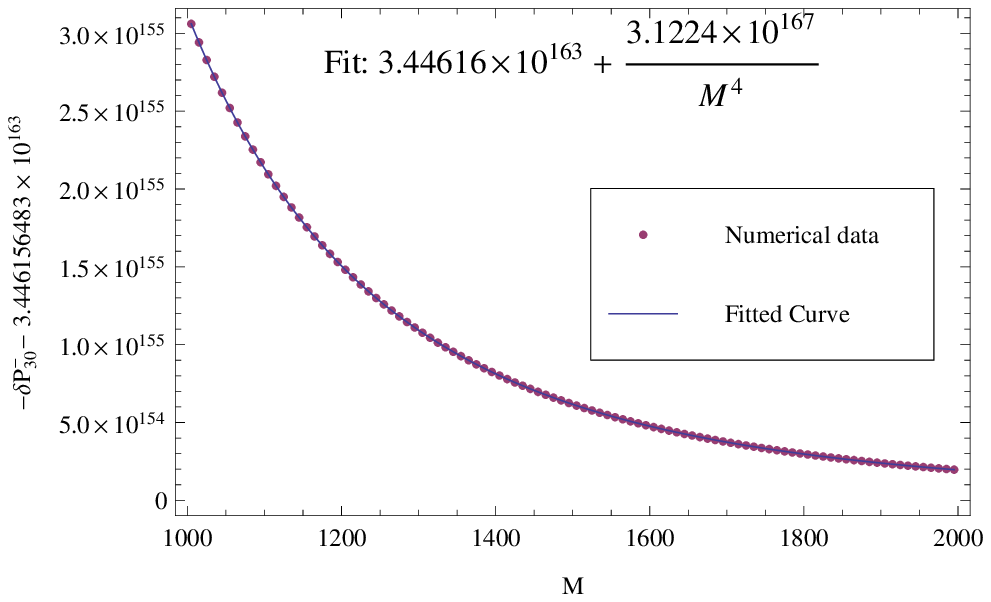}
\caption{\label{plot_tachyon_k130}}
\end{subfigure}
\caption{Plots of tachyon self-energy for an interaction lasting $K-1$ time steps, for $K=14$ in (a) and $K=30$ in (b). We have shifted the vertical axis by a constant so as to depict the much smaller variation of $\delta P^-_K$ more clearly. The fits suggest that it is the $\mathcal{O}(M^{-4})$ term which is responsible for the deviation from the $\mathcal{O}(M)$ behavior.\label{plot_tachyon_k14-30}}
\end{figure}

This can be seen in more detail in Figure \ref{plot_tachyon_k14-30}, where we compare $\delta P^-_K$ against a fit with a constant and an $\mathcal{O}(M^{-4})$ term for $K=14,30$, finding very good agreement. The fit suggests that its two parameters always have comparable sizes and grow very fast with $K$. We already know from the analysis of Section~\ref{sec_slit_size_dependence} that the $\mathcal{O}(M)$ and $\mathcal{O}(M^{-1})$ coefficients also have comparable sizes (due to the same exponential factor), and to give a measure of comparison, they range between order $10^{56}-10^{58}$ for $K=14$ and $10^{128}-10^{131}$ for $K=30$. This in turn implies that for the range of $M$ we are examining, already at $K=14$ the $\mathcal{O}(M^{-4})$ is one order of magnitude larger than the $\mathcal{O}(M)$ term, and their ratio grows to 24 orders of magnitude for $K=30$.

Given that the $\mathcal{O}(M^{-1})$ term is a few orders of magnitude smaller than the $\mathcal{O}(M)$ term, the considerations of the previous paragraph justify why we don't need to include them in order to obtain good fits for the $\delta P^-_K$ depicted in Figure \ref{plot_tachyon_k14-30}. More importantly, they imply that as $K$ increases, it becomes very challenging to extract the $\mathcal{O}(M^{-1})$ dependence by purely numerical analysis. Taking the first difference in $M$ does not improve the resolution substantially, as it removes the large $\mathcal{O}(1)$, but not the $\mathcal{O}(M^{-4})$ term. Hence the only remaining possibilities are to either choose a range of much higher values of $M$ so that the two sets of terms become comparable in size, or drastically increase the precision of the numerics, so as to be able to resolve their difference in size. However since the two sets of terms have different exponential behaviors in $K$, employing any of the two aforementioned options is also expected to increase computation time exponentially.
\begin{figure}
\centering
\includegraphics[width=0.7\textwidth]{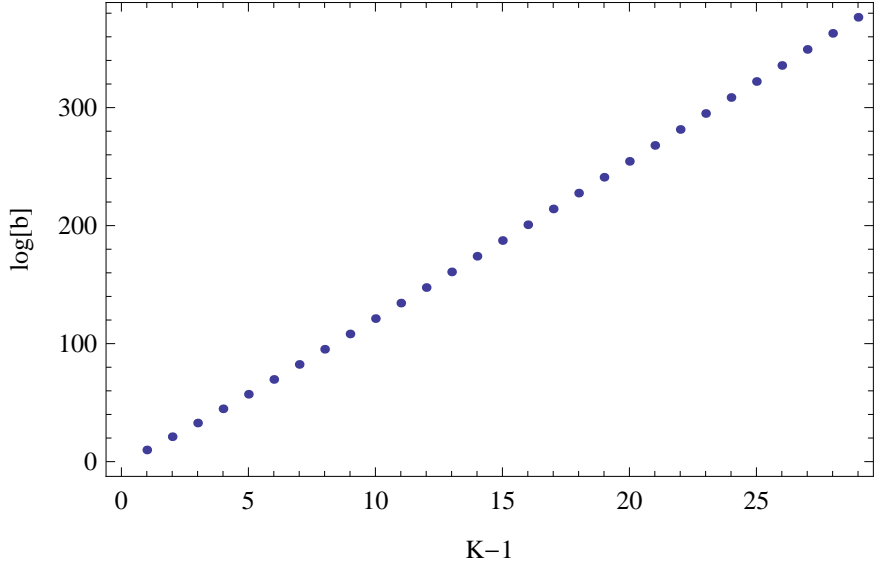}
\caption{Logarithm of the constant term $b$ in the asymptotic expansion in $M$ for the tachyon $\delta P^-_K$ as a function of the interaction time $K-1$. The leading behavior is clearly linear, from which we can infer that $b\propto e^{13K}$.\label{fig:boundary_term_k}}
\end{figure}

In more detail, we can verify that the $\mathcal{O}(1)$ term has an exponential behavior in $K$ of roughly $e^{13K}$ by plotting the logarithm of its fitted value against $K\in[2,30]$, see Figure \ref{fig:boundary_term_k}. As we discuss in the Introduction, the exponential increase in $K$ of the $\mathcal{O}(1)$ and $\mathcal{O}(M^{-4})$ terms is due to the tachyonic divergence, which appears when one of the two intermediate strings becomes very short, namely it is a boundary effect. On the contrary, in the regime $1\ll K\ll M$ we are examining, the tachyonic divergence does not affect the $\mathcal{O}(M)$ and $\mathcal{O}(M^{-1})$ terms, whose exponential dependence $e^{12\beta_0 K}\simeq e^{10.6K}$ is precisely cancelled by the tree-level boundary counterterm.

\begin{figure}
\centering
\begin{subfigure}[b]{0.49\textwidth}
\includegraphics[width=\textwidth]{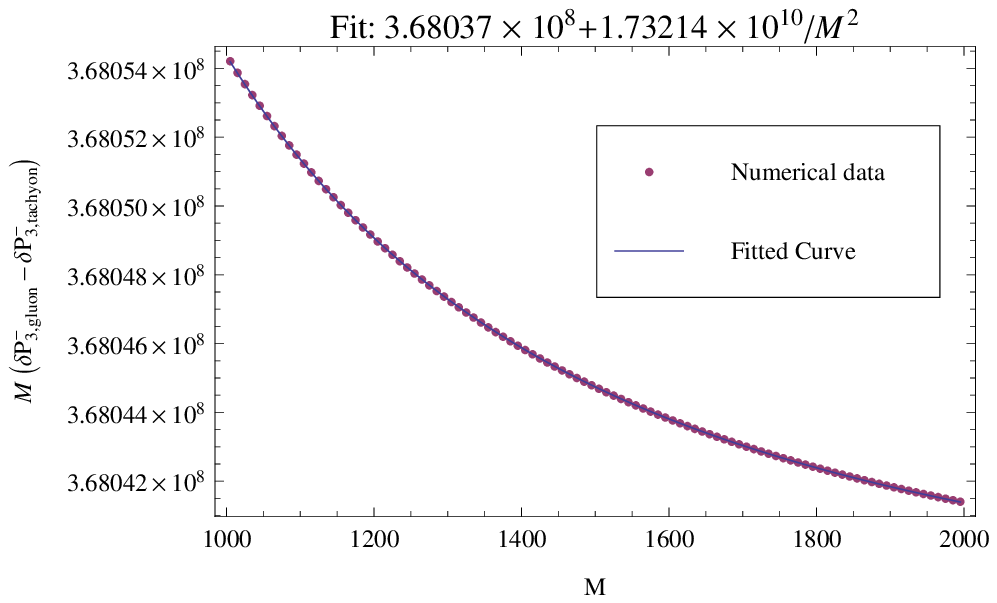}
\caption{\label{plot_graviton_k14}}
\end{subfigure}
\begin{subfigure}[b]{0.49 \textwidth}
\includegraphics[width=\textwidth]{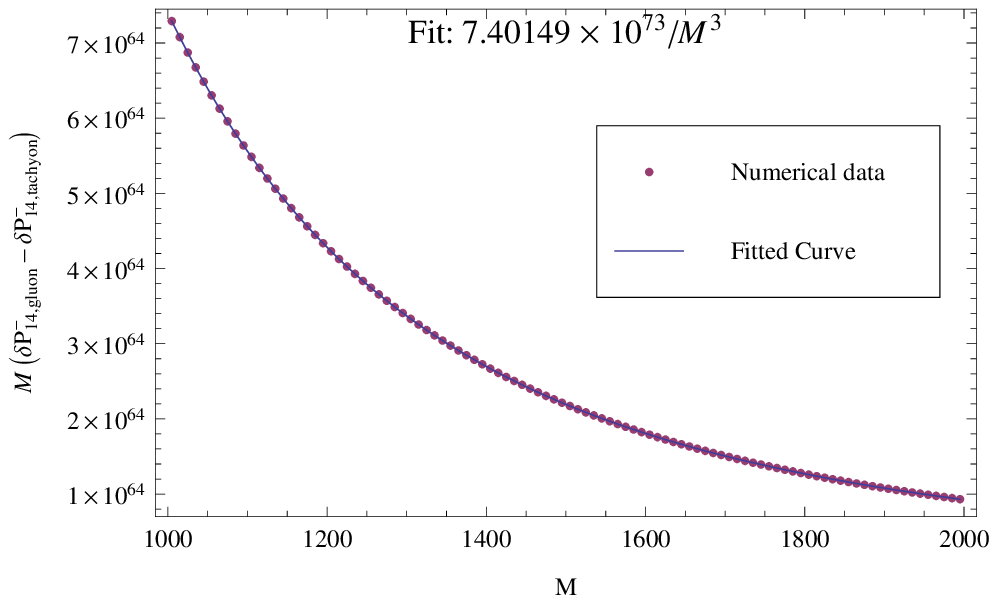}
\caption{\label{plot_graviton_k130}}
\end{subfigure}
\caption{Plots of the self-energy difference between the gluon and the tachyon for $K=3$ in (a) and $K=14$ in (b). We have rescaled the self-energy by $M$ so as to have only one independent variable. As with the tachyon, subleading terms in the $M$ expansion become dominant with increasing $K$.\label{plot_graviton_k3-14}}
\end{figure}

\begin{table}\centering
\begin{tabular}{|c|c|}\hline
$K$&$\delta P^-_{K,\rm{gluon}}-\delta P^-_{K,\rm{tachyon}} \textrm{ fit}$\\\hline
 2 & $-12868.96/M-2.0\times 10^5/M^3 $\\
 3 & $-3.68037\times 10^8/M-1.7\times 10^{10}/M^3$ \\
 4 & $-1.09501\times 10^{13}/M-1.1\times 10^{15}/M^3$\\
 5 & $-3.4332\times 10^{17}/M-7\times 10^{19}/M^3$ \\\hline
\end{tabular}
\caption{Fit of difference of tachyon and gluon self-energy for an interaction lasting $K-1$ time steps (error estimates at the order of the last digit). Comparing with the last row of Table \ref{table_deltaP_comparison}, we see that the $\mathcal{O}(M^{-1})$ terms are identical, thereby supporting that the corresponding term is zero for $\delta P^-_{K,\rm{gluon}}$.\label{table_deltaP_gluon}}
\end{table}
Moving on to a numerical evaluation of the gluon self-energy, we shall aim to compare our fits with the corresponding asymptotic formula, eq. (\ref{deltaP_single_slit}) with $m=1$. In particular we will investigate whether our numerical analysis agrees with the $\mathcal{O}(M^{-1})$ term vanishing, and for that reason it will be more advantageous to examine the quantity
\be\label{deltaP_difference}
\delta P^-_{K,\textrm{\footnotesize gluon}}-\delta P^-_{K,\textrm{\footnotesize tachyon}}=-4\sin^2\frac{\pi}{2M}\sum_{{M_1}=1}^{M-1}\sin^2{ \pi{M_1}\over M}\sum_{q,s=1}^{K-1}e^{(q-s) \lambda^o_{1}} h^{-1}_{qs}\,,
\ee
which has the large $\mathcal{O}(M)$ and $\mathcal{O}(1)$ dependence removed, thereby providing more accurate fits. As for the tachyon, we choose  $K\in{[2,30]}$ and $M\in{[495,1995]}$, and for the first few values of $K$, the fits we obtain are depicted in Table \ref{table_deltaP_gluon}, see also Figure \ref{plot_graviton_k3-14}a.

For small $K$, the numerics suggest that the $\mathcal{O}(M^{-1})$ term is identical here and for the tachyon, implying it should be zero for the gluon. Furthermore, the numerics suggest that the subleading term in the asymptotic expansion of (\ref{deltaP_difference}) is $\mathcal{O}(M^{-3})$. Similarly to the tachyon case however, as $K$ increases the $\mathcal{O}(M^{-4})$ term dominates the expansion to an extent that does not allow the extraction of the $\mathcal{O}(M^{-1})$ dependence by numerical means (see Figure \ref{plot_graviton_k3-14}b). Finally, the $\mathcal{O}(M^{-4})$ term in (\ref{deltaP_difference}) appears to be different from the corresponding term for the tachyon alone, in particular about an order of magnitude larger.

\section{Open Strings Ending on D-branes}\label{sec_Dbranes}
So far we have assumed that all open string transverse coordinates
obey Neumann conditions. But it is also interesting to impose Dirichlet
conditions on a subset of the coordinates \cite{dailp},
denoted by ${\bfs y}(\sigma,\tau)$
to distinguish them from the coordinates ${\bfs x}$ which continue to
satisfy Neumann boundary conditions\footnote{We do not consider here
the mixed case of Neumann and Dirichlet conditions on opposite
ends of an open string.}.
If ${\bfs x}$ has $p-1$ components one says that there is a D$p$-brane at
the location specified by the (fixed) value of ${\bfs y}$ at the
end of each open string. Here we restrict attention to a single
D$p$-brane location at ${\bfs y}=0$. In the continuum, the expressions
for the one loop self-energy of an open string with $25-p$ Dirichlet
coordinates differs from the expressions (\ref{conttachse}) and
(\ref{contgluonse}) simply by an insertion of the factors
$(-2\pi/\ln q)^{(25-p)/2}$ in the integrands. While these factors
marginally soften the leading UV divergence from the integration
range near $q\sim0$, they
do so in a way that introduces a logarithmic branch point at
$q=0$\footnote{In the  open string nonplanar one-loop
diagram, which contains singularities in the pomeron channel
invariant $t$ due to closed string states,
this branch point in $q$ causes the ``unitarity violating''
branch point (instead of a pole) in $t$ that led Lovelace to
anticipate the need for
the critical dimension $D=26$ \cite{lovelace}. Here we see that
in addition to $D=26$ we also need Neumann boundary conditions
on all string coordinates to cancel the cut. As clarified in
\cite{thornsub}, the branch point is not really
``unitarity violating'', but rather simply a reflection
of a continuous closed string mass spectrum, or the holographic
emergence of extra dimensions for the propagation of closed strings.
In any case the branch point in $t$ also spells difficulty for the
Goddard-Neveu-Scherk analytic continuation method \cite{GNS} of
regulating these divergences.}.
After discretization, the presence of these factors leads
to an expected leading large $M$ behavior
\bea
\delta P^-_K&\sim& {\alpha M\over K^3(\ln(M/K))^{(25-p)/2}}
\eea
which can no longer be cancelled by the bulk worldsheet
cosmological constant. In particular the leading singularity
must be accepted as a real divergence, at least in
the perturbative loop expansion of bosonic string theory\footnote{
Supersymmetry can potentially mitigate these difficulties
through cancellation of the divergences due
to tachyonic closed string states. In such models the UV divergence
due to the dilaton is rendered finite provided that  more
than 2 coordinates are Dirichlet: $\int dq q^{-1}(-\ln q)^{-3/2}$ is
convergent at $q\sim0$ \cite{rojast}.
}. The worldsheet lattice provides a physical cutoff, but there is
no consistent way to define a finite continuum limit in perturbation
theory.
\subsection{D-branes and the GT lattice}
We turn to a detailed analysis of the self-energy on the GT lattice
which will confirm these expectations.
The Dirichlet worldsheet propagator
on the free string worldsheet takes either of the forms
(\ref{Dopenpropfsmodes}) or (\ref{Dopenpropftmodes}). The main
new feature of these formulas is the absence of zero modes
in the open string spectrum
because the Dirichlet conditions break translation invariance.
Of course on the lattice loop corrections will involve a different
choice for the matrix $V$ describing  the breaking and joining
of strings, which we shall denote as $\tilde V$ for clarity. A broken Dirichlet string coordinate $y$ involves
the replacement \cite{thorndirichlet}:
\bea
(y_{k+1}^j-y_k^j)^2+(y_k^j-y_{k-1}^j)^2&\to&(y_{k+1}^j)^2+(y_{k-1}^j)^2
+2\kappa (y_k^j)^2.
\eea
where the parameter $\kappa$ gives us some flexibility in
specifying the Dirichlet condition on the lattice. The matrix $\tilde V$
that describes this replacement is then
\bea
\tilde V_{ml,m^\prime l^\prime}&=&\delta_{lj}\delta_{l^\prime j}
(\delta_{m,k+1}\delta_{m^\prime,k}+\delta_{m,k}\delta_{m^\prime,k+1}
+\delta_{m,k-1}\delta_{m^\prime,k}+\delta_{m,k}\delta_{m^\prime,k-1}\nonumber\\
&&\qquad +2(\kappa-1)\delta_{m,k}\delta_{m^\prime,k})
\eea
In contrast to a broken Neumann coordinate, which involves two
lattice sites, the broken Dirichlet coordinate involves 3
lattice sites: $k-1$, $k$, and $k+1$. In matrix form it is
\bea
\tilde V&=&\pmatrix{0&1&0\cr 1&2(\kappa-1)&1\cr 0&1&0\cr}\label{V_Dirichlet}
\eea
Writing the corresponding $3\times3$ block of the propagator as
\bea
\Delta&=&\pmatrix{e&a&f\cr a&c&b\cr f&b&d\cr}
\eea
we find after a simple calculation
\bea
\det(1+\tilde V\Delta)&=&(1+a+b)^2-c(e+d+2f-2(\kappa-1))
\eea
The matrix elements are related to the worldsheet propagator as
follows:
\bea
a&=&\Delta_{kj,(k-1)j}=\Delta_{(k-1)j,kj},\qquad b=\Delta_{kj,(k+1)j}
=\Delta_{(k+1)j,kj}\\
c&=&\Delta_{kj,kj},\qquad d=\Delta_{(k+1)j,(k+1)j},
\qquad e=\Delta_{(k-1)j,(k-1)j}\\
f&=&\Delta_{(k+1)j,(k-1)j}=\Delta_{(k-1)j,(k+1)j}
\eea
We will now proceed to evaluate the contribution of Dirichlet coordinates
to the one loop self-energy calculation with $K=2$. As we did in the previous sections, we will again relabel $k\to M_1$ to better convey that it is referring to the position of the slit. We start by using the representation (\ref{Dopenproptmodes}) for the Dirichlet worldsheet propagator in order to obtain
\bea
c&=&\int_0^1dx{\sinh {M_1}\lambda\sinh(M-{M_1})\lambda
\over\sinh M\lambda\sinh\lambda}\\
a&=&\int_0^1dx{\sinh({M_1}-1)\lambda\sinh(M-{M_1})\lambda
\over\sinh M\lambda\sinh\lambda}\nonumber\\
&=&\int_0^1dx{\sinh {M_1}\lambda\sinh(M-{M_1})\lambda
\over\sinh M\lambda}\coth\lambda-\int_0^1 dx{\cosh {M_1}\lambda\sinh(M-{M_1})\lambda
\over\sinh M\lambda}\\
b&=&\int_0^1dx{\sinh {M_1}\lambda\sinh(M-{M_1})\lambda
\over\sinh M\lambda}\coth\lambda-\int_0^1 dx{\sinh {M_1}\lambda\cosh(M-{M_1})\lambda
\over\sinh M\lambda}
\eea
\bea
1+a+b&=&2\int_0^1dx{\sinh {M_1}\lambda\sinh(M-{M_1})\lambda
\over\sinh M\lambda}\coth\lambda\nonumber\\
&=&2\int_0^1dx{\sinh {M_1}\lambda\sinh(M-{M_1})\lambda
\over\sinh M\lambda}\left({1\over\sinh\lambda}+\tanh{\lambda\over2}\right)\\
&=&2c+2\int_0^1dx{\sinh {M_1}\lambda\sinh(M-{M_1})\lambda
\over\sinh M\lambda}\tanh{\lambda\over2}\nonumber\\
d+e+2f&=&4\int_0^1 dx{\sinh {M_1}\lambda\sinh(M-{M_1})\lambda
\over\sinh M\lambda}\left({1\over\sinh\lambda}+\sinh\lambda\right)
-2\int_0^1dx\cosh\lambda\nonumber\\
&=&4c+4\int_0^1 dx{\sinh {M_1}\lambda\sinh(M-{M_1})\lambda
\over\sinh M\lambda}\sinh\lambda
-2\int_0^1dx\cosh\lambda
\eea
Inserting these into the formula for the
determinant leads to
\bea
\det(I+\tilde V\Delta)&=&c\int_0^1dx\left(4\sinh^2(\lambda/2)
-8{\sinh {M_1}\lambda\sinh(M-{M_1})\lambda
\over\sinh M\lambda}\left[{\sinh^3(\lambda/2)\over\cosh(\lambda/2)}\right]
+2\kappa\right)\nonumber\\
&&+4\left(\int_0^1dx{\sinh {M_1}\lambda\sinh(M-{M_1})\lambda
\over\sinh M\lambda}\tanh{\lambda\over2}\right)^2
\eea
We will require the large $M$ limit of this determinant.
We notice that the quantity squared in the last term is just the
determinant we encountered for Neumann coordinates (\ref{Dk})
\be
2\int_0^1dx{\sinh {M_1}\lambda\sinh(M-{M_1})\lambda
\over\sinh M\lambda}\tanh{\lambda\over2}=
{\rm D}_{M_1}={1\over2}-I_{M_1}+{f_2(x)\over2\pi M^2}+{f_4(x)\over12\pi M^4}
+\cdots
\ee
where we recall (\ref{Dklargem}) and our definition $x={M_1}/M$.

We also have some new integrals to analyze:
\bea
\int_0^1dx\sinh^2{\lambda\over2}&=&{1\over\pi}\int_0^{\lambda_0}d\lambda
{\cosh(\lambda/2)\sinh^2(\lambda/2)\over\sqrt{1-\sinh^2(\lambda/2)}}={1\over2}
\eea
and for ${M_1}\leq M/2$ we use
\bea
{\sinh {M_1}\lambda\sinh(M-{M_1})\lambda
\over\sinh M\lambda}&=& {1\over2}(1-e^{-2{M_1}\lambda})
-{e^{-M\lambda}\sinh^2 {M_1}\lambda
\over\sinh M\lambda},\qquad {M_1}\leq {M\over2}
\eea
to decompose the second integral into two terms
\bea
\int_0^1 dx{1\over2}(1-e^{-2{M_1}\lambda})
{\sinh^3(\lambda/2)\over\cosh(\lambda/2)}
&=&{1\over2\pi}\int_0^{\lambda_0} d\lambda(1-e^{-2{M_1}\lambda})
{\sinh^3(\lambda/2)\over\sqrt{1-\sinh^2(\lambda/2)}}\\
&=&{1\over2\pi}-J_{M_1}\nonumber\\
J_{M_1}&=&{1\over2\pi}\int_0^{\lambda_0} d\lambda e^{-2{M_1}\lambda}
{\sinh^3(\lambda/2)\over\sqrt{1-\sinh^2(\lambda/2)}}\\
\int_0^1 dx{-e^{-M\lambda}\sinh^2 {M_1}\lambda
\over\sinh M\lambda}
{\sinh^3(\lambda/2)\over\cosh(\lambda/2)}
&=&{1\over\pi}\int_0^{\lambda_0} d\lambda{-e^{-M\lambda}\sinh^2 {M_1}\lambda
\over\sinh M\lambda}
{\sinh^3(\lambda/2)\over\sqrt{1-\sinh^2(\lambda/2)}}\nonumber\\
&=&{h_4(x)\over M^4}+{\mathcal O}(M^{-6},e^{-M\lambda_0})\\
h_4(x)&=&-{1\over8\pi}\int_0^\infty \lambda^3d\lambda
{e^{-\lambda}\sinh^2 x\lambda\over\sinh \lambda}\eea
In contrast to the above integrals which are finite as $M\to\infty$,
the integral defining $c$ increases logarithmically with $M$.
This feature is a direct consequence of
Dirichlet boundary conditions. Remembering that the determinant
enters the self-energy with a negative power, this means that
Dirichlet conditions soften the leading UV divergence by powers
of $(\ln M)^{-1}$. To investigate this phenomenon we look at
\bea
c&=&\int_0^{\lambda_0} d\lambda{\coth(\lambda/2)
\over\pi\sqrt{1-\sinh^2(\lambda/2)}}{\sinh {M_1}\lambda\sinh(M-{M_1})\lambda
\over\sinh M\lambda}\equiv G_{M_1}+\hat{c}\\
G_{M_1}&=&{1\over2\pi}\int_0^{\lambda_0}d\lambda
{(1-e^{-2{M_1}\lambda})\coth(\lambda/2)
\over\sqrt{1-\sinh^2(\lambda/2)}}\\
\hat{c}&=&{1\over\pi}\int_0^{\lambda_0} d\lambda
{\coth(\lambda/2)
\over\sqrt{1-\sinh^2(\lambda/2)}}\left[{e^{-M\lambda}\sinh^2 {M_1}\lambda
\over\sinh M\lambda}\right]
\eea
The large $M$ behavior of $\hat{c}$ can be obtained by expanding
\bea
{\coth(\lambda/2)\over\sqrt{1-\sinh^2(\lambda/2)}}
&=&{2\over\lambda}+{5\lambda\over12}+{\mathcal O}(\lambda^3)
\eea
Then term by term we can extend the upper limit to $\infty$, with
errors smaller than  order $e^{-M\lambda_0}$ to get an
asymptotic expansion
\bea
\hat{c}&=&g_0(x)+{g_2(x)\over M^2}+{\mathcal O}(M^{-4},e^{-M\lambda_0})\\
g_0(x)&=&{2\over\pi}\int_0^\infty{d\lambda\over\lambda}
\left[{e^{-\lambda}\sinh^2(x\lambda)
\over\sinh \lambda}\right],\qquad
g_2(x)={5\over 12\pi}\int_0^\infty{\lambda d\lambda}
\left[{e^{-\lambda}\sinh^2(x\lambda)
\over\sinh \lambda}\right]
\eea
In this expansion the coefficients depend on the ratio $x={M_1}/M$ which
is smaller than $1/2$: if this ratio is of order $1/M$, all of
the terms are down by a further factor of $1/M^2$.

Next we study the difference
$G_{M_1}=c-\hat{c}$ which depends only on ${M_1}$. Since ${M_1}$ has
to be summed over the range $0<{M_1}<M/2$, we will need the large ${M_1}$
behavior of this term, which we will find behaves like $\ln {M_1}$.
\bea
G_{M_1}&=&{1\over2\pi}\int_0^{\lambda_0}d\lambda
{(1-e^{-2{M_1}\lambda})\coth(\lambda/2)\over\sqrt{1-\sinh^2(\lambda/2)}}\nonumber\\
&=&{1\over2\pi}\int_0^{\lambda_0}\hspace{-8pt}d\lambda (1-e^{-2{M_1}\lambda})\hspace{-4pt}\left[
{\coth(\lambda/2)\over\sqrt{1-\sinh^2(\lambda/2)}}-{2\over\lambda}\right]
+{1\over\pi}\int_0^{\lambda_0}\hspace{-8pt}d\lambda{1-e^{-2{M_1}\lambda}\over\lambda}
\eea
The quantity in square brackets has good small $\lambda$ dependence so
the terms in the first integral can be evaluated separately:
\bea
{1\over2\pi}\int_0^{\lambda_0}d\lambda \left[
{\coth(\lambda/2)\over\sqrt{1-\sinh^2(\lambda/2)}}-{2\over\lambda}\right]
&=&{1\over\pi}\ln{2\over\lambda_0}+{\Gamma^\prime(1)\over2\pi}-
{\Gamma^\prime(1/2)\over2\pi\sqrt{\pi}}\\
{1\over2\pi}\int_0^{\lambda_0}d\lambda e^{-2{M_1}\lambda}\left[
{\coth(\lambda/2)\over\sqrt{1-\sinh^2(\lambda/2)}}-{2\over\lambda}\right]
&\sim&{5\over24\pi (2{M_1})^2}+{\mathcal O}({M_1}^{-4},e^{-2{M_1}\lambda_0})
\eea
Finally the large ${M_1}$ behavior of the last integral
\bea
{1\over\pi}\int_0^{\lambda_0}\hspace{-8pt}d\lambda{1-e^{-2{M_1}\lambda}\over\lambda}
&=&-{1\over\pi}\int_0^{\lambda_0}\hspace{-8pt}{d\lambda}(2{M_1}
e^{-2{M_1}\lambda})
\ln{\lambda\over\lambda_0} =-{1\over\pi}\int_0^{\infty}\hspace{-8pt}{d\lambda}e^{-\lambda}
\ln{\lambda\over2{M_1}\lambda_0}+{\mathcal O}(e^{-2{M_1}\lambda_0})\nonumber\\
&=&{1\over\pi}\ln(2{M_1}\lambda_0)-{\Gamma^\prime(1)\over\pi}
+{\mathcal O}(e^{-2{M_1}\lambda_0})
\eea
All together then we have for the large ${M_1}$ behavior of $c-\hat{c}$
\bea
G_{M_1}&=&{1\over\pi}\ln(4{M_1})-{\Gamma^\prime(1)
+\Gamma^\prime(1/2)/\sqrt{\pi}\over2\pi}-{5\over96\pi {M_1}^2}
+{\mathcal O}({M_1}^{-4},e^{-2{M_1}\lambda_0})\nonumber\\
&=&{1\over\pi}\ln(4{M_1})-{\psi(1)
+\psi(1/2)\over2\pi}-{5\over96\pi {M_1}^2}
+{\mathcal O}({M_1}^{-4},e^{-2{M_1}\lambda_0})
\eea
where $\psi(z)\equiv\Gamma^\prime(z)/\Gamma(z)$ is the digamma function.

Finally we quote the determinant for $K=2$
due to a Dirichlet coordinate keeping
terms up to order $M^{-2}$:
\bea
\det(I+\tilde V\Delta)&=&\left(G_{M_1}+g_0(x)+{g_2(x)\over M^2}\right)\left(
2(\kappa+1)-{4\over\pi}+8J_{M_1}\right)\nonumber\\
&&+\left({1\over2}-I_{M_1}\right)^2+{(1-2I_{M_1})f_2(x)\over2\pi M^2}
+{\mathcal O}(M^{-4})\,.
\eea
Then the self-energy shift of the tachyon in the presence of
a D$p$-brane is given by
\bea
\delta P^-_{K}&=&2g^2\sum_{M_1=1}^{(M-1)/2}{\det}^{-(p-1)/2}(I+V\Delta)
{\det}^{-(25-p)/2}(I+\tilde{V}\Delta)e^{-24B(K-1)}
\eea
The leading behavior for large $M$ occurs from the region
$1\ll M_1={\mathcal O}(M)$. For $K=2$ this gives
\bea
\delta P^-_{K=2}&\sim&2g^2\sum_{M_1\gg1}^{(M-1)/2}2^{(p-1)/2}
\left({2(\kappa+1)-4/\pi)\over\pi}\ln M_1\right)^{-(25-p)/2}e^{-24B}
\nonumber\\
&\sim&g^22^{p-13}{M\over(\ln M)^{(25-p)/2}}
\left({\pi^2\over(\kappa+1)\pi-2}\right)^{(25-p)/2}e^{-24B}
\eea
Subleading divergences of the form $M/(\ln M)^{n+(25-p)/2}$
and  $1/(\ln M)^{n+(25-p)/2}$ will also appear. In fact,
each power of $M$ can be expected to be multiplied by a power
series in $(\ln M)^{-1}$. We leave the interpretation of these
non-analytic divergences to future work.

\subsection{Discretization of the
continuum expressions for the self-energy of superstring.}
Although we do not yet have a completely satisfactory GT lattice for the
superstring, we can get a glimpse of the benefits of supersymmetry
by simply discretizing the known continuum formulas for the gluon self-energy
diagrams for the superstring with supersymmetry broken
by compactification of an extra dimension\footnote{With unbroken supersymmetry
the diagram is identically zero!}.
The critical dimension is $D=10$ so there
will be $8$ transverse coordinates ${\bfs x},{\bfs{\cal P}}$ and 8 worldsheet
fermions denoted ${\bfs \Gamma}$ in the Ramond (R) sector and ${\bfs H}$
in the Neveu-Schwarz (NS) sector. We shall need the correlators:
\bea
\VEV{{{\cal P}}{{\cal P}}}&=&{1\over4\sin^2(\theta/2)}
-\sum_{n=1}^\infty{2nq^{2n}\over1-q^{2n}}\cos n\theta\\
&=&{1\over4\sin^2(\theta/2)}
-{2q^{2}}\cos \theta+{\mathcal O}(q^4)\\
\VEV{HH}_+&=&{1\over2\sin(\theta/2)}-2\sum_{r=1/2}^\infty{q^{2r}\over1+q^{2r}}
\sin r\theta\\
&=&{1\over2\sin(\theta/2)}-2q\left(1-q+
4q^2\cos^2{\theta\over2}\right)\sin{\theta\over2}
+{\mathcal O}(q^4)\\
\VEV{HH}_-&=&{1\over2\tan(\theta/2)}-2\sum_{n=1}^\infty{q^{2n}\over1+q^{2n}}
\sin n\theta\\
&=&{1\over2\tan(\theta/2)}-2{q^{2}}
\sin\theta+{\mathcal O}(q^4)\\
\VEV{\Gamma\Gamma}&=&{1\over2\sin(\theta/2)}
+2\sum_{r=1/2}^\infty{q^{2r}\over1-q^{2r}}\sin r\theta\\
&=&{1\over2\sin(\theta/2)}+2q\left(1+q+4q^2\cos^2{\theta\over2}\right)
\sin{\theta\over2}
+{\mathcal O}(q^4)
\eea
where we use the notation and conventions of \cite{thornsubqcd}. We have
expressed the correlators in terms of the moduli $q,\theta$ of the cylinder.
To break supersymmetry, we compactify one of the
transverse target space dimensions imposing periodic boundary conditions
on bosonic states and antiperiodic conditions on fermionic states.
In the expressions of the one loop self-energies, this simply means
an insertion of the factor
\bea
&&\sum_{m=-\infty}^\infty q^{m^2R^2T_0/4\pi^2},\ {\rm bosonic~loop},
\qquad\sum_{m=-\infty}^\infty(-)^m q^{m^2R^2T_0/4\pi^2},
\ {\rm fermionic~loop}.
\eea
Both these factors approach unity in the decompactification limit
$R\to\infty$. The absence of tachyonic divergences in the loop
integrals requires $R^2T_0\geq4\pi^2$. When the gluon polarization
is in the direction of a compactified coordinate, the $\VEV{{\cal P}{\cal P}}$
correlator quoted above acquires an extra term:
\bea
\VEV{{{\cal P}}{{\cal P}}}&=&-\left\langle{{m^2R^2T_0\over4\pi^2}}\right\rangle
+{1\over4\sin^2(\theta/2)}
-\sum_{n=1}^\infty{2nq^{2n}\over1-q^{2n}}\cos n\theta
\label{extraterm}\eea
where
\bea
\left\langle{f(m)}\right\rangle&\equiv&
{\sum_m f(m)q^{m^2R^2T_0/4\pi^2}\over
\sum_m q^{m^2R^2T_0/4\pi^2}},\qquad{\rm bosonic~loop}\\
\left\langle{f(m)}\right\rangle&\equiv&
{\sum_m (-)^mf(m)q^{m^2R^2T_0/4\pi^2}\over
\sum_m (-)^mq^{m^2R^2T_0/4\pi^2}},\qquad{\rm fermionic~loop}
\eea
By taking the $R\to0$ limit we get the extra term in this correlator
when gluon polarizations
are transverse to a D-brane: it is just $1/(2\ln q)$.
It is noteworthy that the extra term from either compactification
or from the presence of D-branes is negative. Since generally
$C_s<0$,\footnote{From the lightcone viewpoint the self-energy shift
is a result from second order perturbation theory which is necessarily
negative by unitarity. Since the divergence in the integral has a positive
coefficient, this implies that $C_s$ must be negative. The negative
divergent contribution to the shift is, of course,
cancelled against the boundary cosmological constant counterterm $B$.}
this term therefore contributes positively to the self energy.
Thus while the mass shift of the gluon (polarizations in uncompactified
directions) is zero, the mass squared shift of the massless scalar
(polarization in compactified directions) is positive.

The self-energy of the gluon state is given as a sum of three terms
\bea
\Delta P^-&=&{C_s\over2P^+}\left(\Sigma_++\Sigma_-+\Sigma_F\right)\\
\Sigma_+&=&{1\over2}\int_0^1{dq\over q^2}\int_0^{2\pi}d\theta
\sum_{m=-\infty}^\infty q^{m^2R^2T_0/4\pi^2}
{\prod_r(1+q^{2r})^8\over\prod_n(1-q^{2n})^8}
\VEV{{{\cal P}}{{\cal P}}}\\
\Sigma_-&=&-8\int_0^1{dq\over q^2}\int_0^{2\pi}d\theta
\sum_{m=-\infty}^\infty q^{m^2R^2T_0/4\pi^2}
{\prod_n(1+q^{2n})^8\over\prod_n(1-q^{2n})^8}
q\VEV{{{\cal P}}{{\cal P}}}\\
\Sigma_F&=&-{1\over2}\int_0^1{dq\over q^2}\int_0^{2\pi}d\theta
\sum_{m=-\infty}^\infty (-)^mq^{m^2R^2T_0/4\pi^2}
{\prod_r(1-q^{2r})^8\over\prod_n(1-q^{2n})^8}
\VEV{{{\cal P}}{{\cal P}}}
\eea
Terms involving the fermionic correlators
$\VEV{HH}^2_\pm$ and $\VEV{\Gamma\Gamma}^2$ do not contribute to the onshell
two gluon function because they are multiplied by kinematic
factors like $k_i\cdot k_j \epsilon_k\cdot\epsilon_l$ or
$k_i\cdot\epsilon_j k_k\cdot\epsilon_l$. But for the two point
function $k_2=-k_1$, $k_i^2=0$, and $k_i\cdot\epsilon_i=0$, so
all these factors vanish.
The combination $\Sigma_++\Sigma_-$ projects out the odd G-parity states
of the NS sector circulating the loop, while $\Sigma_F$ represents
the R sector states circulating the loop.
Because of Jacobi's abstruse identity
\bea
\prod_r(1+q^{2r})^8-\prod_r(1-q^{2r})^8
-16q\prod_n(1+q^{2n})^8&=&0
\eea
We have, for gluon polarization in uncompactified directions,
 the simplification
\bea
\Delta P^-&=&{C_s\over2P^+}\int_0^1{dq\over q^2}\int_0^{2\pi}d\theta
\sum_{m={\rm odd}}q^{m^2R^2T_0/4\pi^2}
{\prod_r(1-q^{2r})^8\over\prod_n(1-q^{2n})^8}\VEV{{{\cal P}}{{\cal P}}}
\label{supergluonse}
\\
&&\hskip-1in={C_s\over2P^+}\int{dq\over q}\int_0^{2\pi}d\theta
 \sum_{m={\rm odd}}q^{m^2R^2T_0/4\pi^2}
\left({1-8q+36q^2\over4q\sin^2(\theta/2)}-2q+4q\sin^2{\theta\over2}
+{\mathcal O}(q^2)\right)
\eea
where in the second line we have explicitly displayed the first few
terms in the $q$ expansion.
It is worth emphasizing that the right side vanishes in the
decompactification limit $R\to\infty$, which restores
supersymmetry. Also notice that,
although the $q$ integral is convergent at the lower end $q\sim0$
when $R^2T_0>4\pi^2$, the $\theta$ integral is still divergent
at its end points. It is this divergence that discretization
will show can be absorbed in the constant boundary counterterm $B$.

Incidentally, for gluon polarization in the compactified direction,
the open string state corresponds to a massless scalar particle which gains 
a mass by virtue of the extra term shown in (\ref{extraterm}):
\bea
\Delta M_{\rm Scalar}^2&=&2P^+\Delta P^-=-{C_sR^2T_0\over2\pi}
\int_0^1{dq\over q^2}
{\prod_r(1-q^{2r})^8\over\prod_n(1-q^{2n})^8}
\sum_{m={\rm odd}}{m^2}q^{m^2R^2T_0/4\pi^2}
\label{superscalarse}
\eea 
which is positive since $C_s<0$.
The integral on the right is convergent at $q\sim0$ provided $R^2T_0>4\pi^2$,
and becomes arbitrarily large as $R^2T_0\to4\pi^2$. The convergence of the
integral at $q\sim1$ becomes transparent after the change of variables
by the Jacobi imaginary transformation $q=e^{2\pi^2/\ln w}$, which maps
$q=1$ to $w=0$.  

Now let's discretize the gluon self-energy (\ref{supergluonse}) 
in the variables of the lightcone
lattice and examine how the continuum limit is regained. Recalling
(\ref{qtot}), (\ref{qtotdiscrete}), and (\ref{qtotjacobian}), we have
\bea
\Delta P^-&\to&{ 64C_s\over a\pi T_0}
\sum_{K}{2+{\mathcal O}(K^4)\over K^3}\sum_{M_1=1}^{(M-1)/2}
\sum_{k=-\infty}^\infty q^{1+(2k-1)^2R^2T_0/4\pi^2}\nonumber\\
&&\qquad\qquad \left[1-8q+36q^2+4q^2\sin^4{\pi M_1\over M}
-2q^2\sin^2{\pi M_1\over M}+{\mathcal O}(q^3)\right]
\eea
where, to avoid ungainly expressions,
we have deferred replacing $q$ with its
discretized version given by (\ref{qtotdiscrete}). Instead we
work out the large $M$ limit of the contribution of each power
of $q$ in what follows.

As we did in subsection \ref{subsection_continuous_selfenergy}, we next study the large $M$ limit
of the terms at fixed $K$.
The summand involves terms of the form
$q^p\sin^{2n}(\pi M_1/M)$ with $p\geq 1+R^2T_0/(4\pi^2)\geq2$ and $n=0,1,2$.
For simplicity of discussion, let's choose $R^2T_0=4\pi^2$, so the
lowest power is $q^2$, and begin by examining the large $M$ limit of
\bea
\sum_{M_1=1}^{(M-1)/2} q^2&\sim&\left(1+{\pi^2K^2\over48M^2}\right)
\sum_{M_1=1}^{(M-1)/2}\left({\pi K\over8M\sin\pi M_1/M}\right)^2-4
\sum_{M_1=1}^{(M-1)/2}
\left({\pi K\over8M\sin\pi M_1/M}\right)^4\nonumber\\
&=&\left(1+{\pi^2K^2\over48M^2}\right)\left({\pi^2 K^2\over384}+{\mathcal O}(M^{-2})\right)-4\zeta(4){K^4\over8^4}+{\mathcal O}(M^{-2})\nonumber\\
&=&{\zeta(2)K^2\over64}-{\zeta(4)K^4\over1024}+{\mathcal O}(M^{-2})
\eea
where the last 2 lines follow from the argument in footnote
\ref{footnote_csc_sums}. A similar analysis applies to the higher powers
of $q$ as well. As this $q^2$ example shows the presence of
additional factors of $\sin(\pi M_1/M)$ have the effect of
suppressing the contribution by additional powers of $M^{-1}$.
This means that in collecting the
contributions to the constant boundary counterterm, one only needs to keep
the first three terms in the square brackets:
\bea
\sum_{M_1=1}^{(M-1)/2} q^3&\sim&
\sum_{M_1=1}^{(M-1)/2}\left({\pi K\over8M\sin\pi M_1/M}\right)^3\nonumber\\
&\sim&{\pi^3K^3\over 8^3M^2}\int_0^{1/2}dx\left({1\over\sin^3\pi x}-{1\over\pi^3 x^3}-{1\over2\pi x}\right)\nonumber\\
&&\qquad+{K^3\over 8^3}\sum_{M_1=1}^{(M-1)/2}{1\over M_1^3}+{\pi^2K^3\over 8^3M^2}
\sum_{M_1=1}^{(M-1)/2}{1\over2M_1}\nonumber\\
&=&{\zeta(3)K^3\over8^3}+{\mathcal O}(M^{-2}\ln M)\\
\sum_{M_1=1}^{(M-1)/2} q^4&\sim&
\sum_{M_1=1}^{(M-1)/2}\left({\pi K\over8M\sin\pi M_1/M}\right)^4
={\zeta(4)K^4\over8^4}+{\mathcal O}(M^{-2})
\eea
Thus the total contribution from these terms to the divergence is
\bea
\Delta P^-_{\rm div}&=&{C_s\over a\pi T_0}
\sum_{K}{1\over K}\left[2{\zeta(2)}-2{\zeta(3)K}+{\zeta(4)K^2}
+{\mathcal O}(K^4)\right]
\eea

\subsection{D-branes and the superstring}
We now return to the effect of imposing Dirichlet conditions on
some of the coordinates. As before, the self-energy integrands
now acquire extra factors $(-\ln q)^{-(9-p)}$, which become
\bea
\left(-\ln{\pi K\over8M\sin\pi M_1/M}+{6-2\sin^2\pi M_1/M\over3}
\left({\pi K\over8M\sin\pi M_1/M}\right)^2+{\mathcal O}(K^4)\right)^{-(9-p)}
\eea
After expanding in powers of $K$, we see that in general we will encounter,
in addition to more terms in higher powers of $K$,
a series of terms with more negative powers of $\ln K$ of the form
\bea
\left(-\ln{\pi K\over8M\sin\pi M_1/M}\right)^{-(9-p+n)}
\eea
These extra factors modify the extraction of the divergent parts
of the discretized self-energy expressions. It will suffice to illustrate this
with the lowest contributing power of $K$:
\bea
\sum_{M_1=1}^{(M-1)/2} q^2(-\ln q)^{-n}&\to&
\sum_{M_1=1}^{(M-1)/2}\left({\pi K\over8M\sin\pi M_1/M}\right)^2
\left(-\ln{\pi K\over8M\sin\pi M_1/M}\right)^{-n}\nonumber\\
&&\hskip-1.5in \sim{\pi^2 K^2\over64M}\int_0^{1/2}dx\left(
{1\over\sin^2\pi x}\left(-\ln{\pi K\over8M\sin\pi x}\right)^{-n}
-{1\over\pi^2 x^2}\left(-\ln{K\over8Mx}\right)^{-n}\right)\nonumber\\
&&\hskip-.5in+\sum_{M_1=1}^{\infty}{K^2\over64M_1^2}\left(-\ln{K\over8M_1}\right)^{-n}-
\sum_{M_1=(M+1)/2}^{\infty}{K^2\over64M_1^2}
\left(-\ln{K\over8M_1}\right)^{-n}\nonumber\\
&=&\sum_{M_1=1}^{\infty}{K^2\over64M_1^2}\left(-\ln{K\over8M_1}\right)^{-n}
+{\mathcal O}(M^{-1}(\ln M)^{-n-1})
\eea
So we see that the presence of D-branes for the superstring does not
spoil the ability to aborb divergences in the boundary counterterm.
Also notice that the negative powers of $\ln M$ suppress
the $M^{-1}$ correction term which, if nonzero, would signal a
nonzero shift to the gluon mass.

\section*{Acknowledgments}
This research was supported in part by the Department
of Energy under Grant No. DE-FG02-97ER-41029.

\appendix
\section{Closed Strings in the Presence of D-branes}

In this appendix, we will briefly examine how the scattering of a closed string tachyon off a D-brane can be described within the worldsheet-based approach, and compare with the string field theory analysis of the same process, which was carried out in Section 3 of \cite{papathanasiout}.

To this end, we will first have to generalize the introductory remarks of Section \ref{sec_Dbranes}, and derive a determinant formula for the path integral where instead of one site obeying Dirichlet boundary conditions, we now have $K-1$ consecutive sites in the temporal direction, and in the same spatial position $k$. This requires taking the direct product of matrix (\ref{V_Dirichlet}) with a diagonal matrix with entries 1 for the sites in question, and zero otherwise. Then $\det(I+\tilde V\Delta)$ may be written in the block form
\[
\det(I+\tilde V\Delta)=\left|
\begin{array}{ccc}
 I+\mathbf{\Delta}_{-10} & \mathbf{\Delta}_{00} & \mathbf{\Delta}_{01} \\
\scriptstyle \mathbf{\Delta}_{-1,-1}+2 (\kappa-1)  \mathbf{\Delta}_{-10}+\mathbf{\Delta}_{-11} &\scriptstyle I+\mathbf{\Delta}_{-10}+2 (\kappa-1)  \mathbf{\Delta}_{00}+\mathbf{\Delta}_{01} &\scriptstyle \mathbf{\Delta}_{-11}+2 (\kappa-1)  \mathbf{\Delta}_{01}+\mathbf{\Delta}_{11} \\
 \mathbf{\Delta}_{-10} & \mathbf{\Delta}_{00} & I+\mathbf{\Delta}_{01}
\end{array}
\right|\,,\]
where each of the blocks corresponds again to spatial positions $k-1,k$ and $k+1$, and $\mathbf{\Delta}_{l m}$ is the $(K-1)$-dimensional matrix with elements $\Delta_{i(k+l),j(k+m)}$, namely with only the temporal indices $i,j$ varying. By elementary row and column operations, we can reduce the size of the matrix and bring it to the form
\bea
\det(I+\tilde V\Delta)&=&\left|\matrix{ I+\mathbf{\Delta}_{-10}+\mathbf{\Delta}_{01}&-2(\kappa-1) I+\mathbf{\Delta}_{-1, -1}+\mathbf{\Delta}_{11}+2\mathbf{\Delta}_{-11}\cr \mathbf{\Delta}_{00}& I+\mathbf{\Delta}_{-10}+\mathbf{\Delta}_{01}}\right|\label{det_Dbrane}\\
&\hspace{-45pt} =&\hspace{-27pt} \left|
\begin{array}{cc}
\scriptstyle I+\mathbf{\Delta}_{-10}-2 \mathbf{\Delta}_{00}+\mathbf{\Delta}_{01} &\scriptstyle -2(\kappa+1)I +\mathbf{\Delta}_{-1,-1}-4 \mathbf{\Delta}_{-10}+2 \mathbf{\Delta}_{-11}+4 \mathbf{\Delta}_{00}-4 \mathbf{\Delta}_{01}+\mathbf{\Delta}_{11} \\
\scriptstyle \mathbf{\Delta}_{00} &\scriptstyle I+\mathbf{\Delta}_{-10}-2 \mathbf{\Delta}_{00}+\mathbf{\Delta}_{01}\nonumber
\end{array}
\right|\,.
\eea
The latter equation may be used for the study of the open or closed string case, by substituting the corresponding expression for the propagator. In what follows we will focus on the closed string, where the propagator is a function of $|l-m|$ alone, and more concretely is given  by (\ref{closed_Delta}). Clearly the zero mode piece of the propagator will dominate as $N\to \infty$, and following the same logic as in the string field theory approach \cite{papathanasiout}, the quantity of interest will be precisely the coefficient of the zero mode in $\det(I+\tilde V\Delta)$,
\be
\mathcal{M}_K=\lim_{N\to\infty}\frac{4M \det(I+\tilde V\Delta)}{N+1}\,.
\ee
The expression of the second line of (\ref{det_Dbrane}) is advantageous for obtaining $\mathcal{M}_K$, as the $\mathcal{O}(N)$ term is contained only in the lower left block. In fact, since this term will be the same for all elements in the block, we can perform further row and columns operations in order to remove it from all but one element. This implies that $\mathcal{M}_K$ will simply equal the minor of the latter element, or in other words is will be given by a determinant of dimension $2(K-1)-1$.

For specific $K$, here it is also possible to obtain and asymptotic expansion in $M$ for $\mathcal{M}_K$ with the help of the Euler-Maclaurin formula. For example, setting $\kappa=1$ for simplicity, we find
\bea
\mathcal{M}_2&=&4(1-\frac{1}{\pi})+\frac{\pi^3}{15}\frac{1}{M^4}+\mathcal{O}(M^{-6})\simeq 2.727+\frac{0.517}{M^2}\,,\\
\mathcal{M}_3&=&\frac{(5 \pi -8+) \left(3 \pi ^2+16\right)}{2 \pi ^3}+\frac{\frac{5 \pi }{2}-4}{M^2}+\mathcal{O}(M^{-4})\simeq 5.669+\frac{3.854}{M^2}\,,\\
\mathcal{M}_4&=&\textstyle - \frac{32 \left(1024-296 \pi -75 \pi ^2+12 \pi ^3\right)}{9 \pi ^4}-\frac{16 \left(2048-720 \pi -129 \pi ^2+36 \pi ^3\right)}{27  \pi ^2M^2}+\mathcal{O}(M^{-4})\\
&\simeq&10.003+\frac{22.270}{M^2}\nonumber\,.
\eea
It's worth noting that particularly for $K=2$, the coefficient of the $\mathcal{O}(M^{-2})$ term is zero. We may compare our results of our current approach to D-branes with the string field theory based approach of \cite{papathanasiout}, by noting that the quantity we defined in equation (87) of the latter paper, equals in our current notations to
\be
r_K=\sqrt{\eta^{-K+1}\frac{1-\eta^{2K}}{1-\eta^2}\frac{\det(h_{lp})}{\mathcal{M}_K}}
\ee
where $\eta=1+\kappa-\sqrt{\kappa(2+\kappa)}$ and $\det(h_{lp})$ is given by equation (33) of \cite{papathanasioutwsprop}. Indeed, we have verified that the two approaches yield the same values for $r_K$ for a wide range of $M$ and $K$, and that the fits we obtained in \cite{papathanasiout} for fixed $K$ and varying $M\gg K$ are in good agreement with the asymptotic expressions we can now obtain analytically. As an illustration, with the current methods we find for $\kappa=1$
\be
r_2=\frac{\sqrt{\pi} }{\sqrt{2 (\pi-1) }}\left(1+\frac{1}{6\pi M^2}\right)+\mathcal{O}(M^{-4})\simeq0.856429 + \frac{0.448425}{M^2}+\mathcal{O}(M^{-4})\,
\ee
which compares very well with the fit on the left hand side of Figure 14 in the latter reference.

\section{Open String Self-energy: String Field Viewpoint}
In this appendix, we give the alternative expression for the
self-energy as a concatenation of open string propagators,
along the lines of \cite{papathanasiout}. For the open string self-energy, depicted in Fig.~\ref{openselattice},
we have a total of $N+K-1$ missing
links, $N$ for the open string ends and $K-1$ for the extra two
ends of the two intermediate strings. Let the external string have
$M$ sites and the two intermediate strings
have $M_1$, $M_2=M-M_1$ sites respectively. At each time $j$ there
will be $M$ coordinates $x_i^j$, $i=1,\ldots,M$. If at time $j$ there
are two open strings, we shall identify $x_i^j$, $i=1,\ldots,M_1$,
with string 1 and $x_i^j$, $i=M_1+1,\ldots, M$ with the second one.
The summand of the self-energy diagram will then depend on $M_1,J,K$,
so we write
\bea
\VEV{N+1,\{x^f\}|0,\{x^i\}}^{open}_{M_1,K,J}&=&
\int dx^K_idx^L_i
\VEV{L,\{x^f\}|0,\{x^L\}}^{open}_M
\VEV{K,\{x_<^L\}|0,\{x_<^K\}}_{M_1}^{open}\nonumber\\
&&\hskip-2in
\VEV{K,\{x_>^L\}|0,\{x_>^K\}}_{M_2}^{open}\VEV{J,\{x^K\}|0,\{x^i\}}_M^{open}
e^{-{T_0}[(x^L_{M_1+1}-x^L_{M_1})^2+(x^K_{M_1+1}-x^K_{M_1})^2]/4}\\
&=&{\cal D}_M^{open}(J){\cal D}_{M_1}^{open}(K)
{\cal D}_{M_2}^{open}(K)
{\cal D}_M^{open}(L)\nonumber\\
&&\int dx^K_idx^L_i
e^{iW+(N+K-1)B_0-{T_0}[(x^L_{M_1+1}-x^L_{M_1})^2
+(x^K_{M_1+1}-x^K_{M_1})^2]/4}\nonumber\eea
where we have introduced the notation $x_<$ for $x_i$, $i=1,\ldots,M_1$ and
$x_>$ for $x_i$, $i={M_1+1},\ldots,M$.

We will again want to change integration variables to  normal modes
of either the single external string or the two intermediate strings as
follows:
\bea
x_i&=&{1\over\sqrt{M}}q_0+\sqrt{2\over M}\sum_{m=1}^{M-1}q_m\cos{m\pi\over M}
\left(i-{1\over2}\right)\\
&=&\cases{\displaystyle{{1\over\sqrt{M_1}}q^{(1)}_0+\sqrt{2\over M_1}
\sum_{m=1}^{M_1-1}q^{(1)}_m\cos{m\pi\over M_1}
\left(i-{1\over2}\right)}&$\quad i=1,\ldots,M_1$\cr
\displaystyle{{1\over\sqrt{M_2}}q^{(2)}_0+\sqrt{2\over M_2}
\sum_{m=1}^{M_2-1}q^{(2)}_m\cos{m\pi\over M_2}
\left(i-M_1-{1\over2}\right)}&$\quad i=M_1+1,\ldots,M$\cr}
\eea
The missing link terms in the exponent involve
\bea
x_{M_1+1}-x_{M_1}
&=&-2\sqrt{2\over M}\sum_{m=1}^{M-1}q_m\sin{mM_1\pi\over M}
\sin{m\pi\over 2M}\\
(x_{M_1+1}-x_{M_1})^2&=&{8\over M}\sum_{m^\prime,m^{\prime\prime}=1}^{M-1}
q_{m^\prime}q_{m^{\prime\prime}}\sin{m^\prime M_1\pi\over M}
\sin{m^{\prime\prime} M_1\pi\over M}
\sin{m^\prime\pi\over 2M}\sin{m^{\prime\prime}\pi\over2M}\eea

It is straightforward to relate the $q^{(1)}_m,q^{(2)}_m$ to the
$q_m$:
\bea
q_0^{(1)}&=&\sqrt{M_1\over M}\ q_0+\sqrt{2\over MM_1}\sum_{m^\prime=1}^{M-1}
 q_{m^\prime}U^{(1)}_{m^\prime 0},\qquad q_m^{(1)}={2\over\sqrt{MM_1}}\sum_{m^\prime=1}^{M-1}
q_{m^\prime}U^{(1)}_{m^\prime m}\\
q_0^{(2)}&=&\sqrt{M_2\over M}\ q_0+\sqrt{2\over MM_2}\sum_{m^\prime=1}^{M-1}
 q_{m^\prime}U^{(2)}_{m^\prime 0},\qquad q_m^{(2)}={2\over\sqrt{MM_2}}\sum_{m^\prime=1}^{M-1}
q_{m^\prime}U^{(2)}_{m^\prime m}
\eea
and we note the identity $q_0^{(1)}\sqrt{M_1}+q_0^{(2)}\sqrt{M_2}
=q_0\sqrt{M}$, as expected from the fact that $q_0/\sqrt{M}$ is the center
of momentum of the open string. The matrices $U^{(1)},U^{(2)}$ are
listed in Appendix~\ref{overlap}.
\subsection{Correction to the open string ground energy}
For the ground state it suffices to set $x^i=x^f=0$, so that
the expression for $iW$ simplifies somewhat:
\bea
iW&\to&-{T_0\over2}\Bigg[{(q^L_{0})^2\over L}+{(q^K_{0})^2\over J}
+\sum_{m=1}^{M-1}\sinh\lambda^o_m
\left((q^L_{m})^2\coth L\lambda^o_m
+(q^K_{m})^2\coth J\lambda^o_m
\right)\nonumber\\
&&\hskip-.6in+{(q^{L,1}_{0}-q^{K,1}_{0})^2\over K}
+\sum_{m=1}^{M_1-1}\sinh\lambda^{o,1}_m
\left([(q^{L,1}_{m})^2+(q^{K,1}_{m})^2]\coth K\lambda^{o,1}_m
-2{q^{K,1}_{m}q^{L,1}_{m}\over\sinh K\lambda^{o,1}_m}\right)\nonumber\\
&&\hskip-.6in +{(q^{L,2}_{0}-q^{K,2}_{0})^2\over K}
+\sum_{m=1}^{M_2-1}\sinh\lambda^{o,2}_m
\left([(q^{L,2}_{m})^2+(q^{K,2}_{m})^2]\coth K\lambda^{o,2}_m
-2{q^{K,2}_{m}q^{L,2}_{m}\over\sinh K\lambda^{o,2}_m}\right)\Bigg]
\eea
where $\lambda_m^{o,1},\lambda_m^{o,2}$ are obtained from $\lambda^o_m$
through the substitutions $M\to M_1,M_2$ respectively.

Finally we eliminate the $q^{(1,2)}_m$ in favor of the $q_m$. Because
$U^{(2)}_{m^\prime0}=-U^{(1)}_{m^\prime0}$, we find that the
zero modes combine nicely
\bea
(q^{L,1}_{0}-q^{K,1}_{0})^2+(q^{L,2}_{0}-q^{K,2}_{0})^2
&=&\nonumber\\
&&\hskip-1.5in(q_0^K-q_0^L)^2+{2\over M_1M_2}\sum_{m^\prime,m^{\prime\prime}=1}^{M-1}
(q_{m^\prime}^K-q_{m^{\prime}}^L)(q_{m^{\prime\prime}}^K
-q_{m^{\prime\prime}}^L)U^{(1)}_{m^\prime0}U^{(1)}_{m^{\prime\prime}0}
\eea
From this we see that the zero modes enter the exponent in the combination
\bea
iW_0&=&-{T_0\over2}\left[{(q^L_{0})^2\over L}+{(q^K_{0})^2\over J}
+{(q_0^K-q_0^L)^2\over K}\right]
\eea
So integrating them out simply implements closure on the zero modes.
\bea
\int dq_0^K dq_0^Le^{iW_0}&=&{2\pi\over T_0}\sqrt{JKL\over J+K+L}
={2\pi\over T_0}\sqrt{JKL\over N+1}.
\eea
The contribution of the nonzero modes to the exponent can be
expressed using the following matrix definitions:
\bea
A^{(1)}_{m^\prime m^{\prime\prime}}&\equiv&{4\over MM_1}
\sum_{m=1}^{M_1-1}U^{(1)}_{m^\prime m}U^{(1)}_{m^{\prime\prime}m}
\sinh\lambda^{o,1}_m\coth K\lambda^{o,1}_m\\
B^{(1)}_{m^\prime m^{\prime\prime}}&\equiv&-{4\over MM_1}
\sum_{m=1}^{M_1-1}U^{(1)}_{m^\prime m}U^{(1)}_{m^{\prime\prime}m}
{\sinh\lambda^{o,1}_m\over\sinh K\lambda^{o,1}_m}\\
A^{(2)}_{m^\prime m^{\prime\prime}}&\equiv&{4\over MM_2}
\sum_{m=1}^{M_2-1}U^{(2)}_{m^\prime m}U^{(2)}_{m^{\prime\prime}m}
\sinh\lambda^{o,2}_m\coth K\lambda^{o,2}_m\\
B^{(2)}_{m^\prime m^{\prime\prime}}&\equiv&-{4\over MM_2}
\sum_{m=1}^{M_2-1}U^{(2)}_{m^\prime m}U^{(2)}_{m^{\prime\prime}m}
{\sinh\lambda^{o,2}_m\over\sinh K\lambda^{o,2}_m}
\eea
Taking the limit $L,J\to\infty$,
we define the nonzero mode contribution to $iW$ plus the
missing link terms as
\bea
iW^\prime&=&-{T_0\over2}\Bigg[
\sum_{m=1}^{M-1}\sinh\lambda^o_m
\left((q^L_{m})^2+(q^K_{m})^2\right)\nonumber\\
&&+{2\over KM_1M_2}\sum_{m^\prime,m^{\prime\prime}=1}^{M-1}
(q_{m^\prime}^K-q_{m^{\prime}}^L)(q_{m^{\prime\prime}}^K
-q_{m^{\prime\prime}}^L)U^{(1)}_{m^\prime0}U^{(1)}_{m^{\prime\prime}0}
\nonumber\\
&&\hskip-.6in
+\sum_{m^\prime,m^{\prime\prime}=1}^{M-1}(q^K_{m^\prime}q^K_{m^{\prime\prime}}
+q^L_{m^\prime}q^L_{m^{\prime\prime}})
(A^{(1)}+A^{(2)})_{m^\prime m^{\prime\prime}}
+2\sum_{m^\prime,m^{\prime\prime}=1}^{M-1}q^K_{m^\prime}
q^L_{m^{\prime\prime}}
(B^{(1)}+B^{(2)})_{m^\prime m^{\prime\prime}}\nonumber\\
&&+{4\over M}
\sum_{m^\prime,m^{\prime\prime}=1}^{M-1}(q^K_{m^\prime}q^K_{m^{\prime\prime}}
+q^L_{m^\prime}q^L_{m^{\prime\prime}})\sin{m^\prime M_1\pi\over M}
\sin{m^{\prime\prime} M_1\pi\over M}
\sin{m^\prime\pi\over 2M}\sin{m^{\prime\prime}\pi\over2M}\Bigg]\\
&\equiv&-{T_0\over2}\left[\sum_{m^\prime,m^{\prime\prime}=1}^{M-1}(q^K_{m^\prime}q^K_{m^{\prime\prime}}
+q^L_{m^\prime}q^L_{m^{\prime\prime}})A_{m^\prime m^{\prime\prime}}
+2q^K_{m^\prime}q^L_{m^{\prime\prime}}B_{m^\prime m^{\prime\prime}}
\right]\eea
where we have defined
\bea
A_{m^\prime m^{\prime\prime}}&\equiv&\delta_{m^\prime m^{\prime\prime}}
\sinh\lambda_{m^\prime}^o+(A^{(1)}+A^{(2)})_{m^\prime m^{\prime\prime}}
+{2\over KM_1M_2}U^{(1)}_{m^\prime0}U^{(1)}_{m^{\prime\prime}0}\nonumber\\
&&\qquad+{4\over M}\sin{m^\prime M_1\pi\over M}
\sin{m^{\prime\prime} M_1\pi\over M}
\sin{m^\prime\pi\over 2M}\sin{m^{\prime\prime}\pi\over2M}\\
B_{m^\prime m^{\prime\prime}}&\equiv&(B^{(1)}
+B^{(2)})_{m^\prime m^{\prime\prime}}
-{2\over KM_1M_2}U^{(1)}_{m^\prime0}U^{(1)}_{m^{\prime\prime}0}
\eea
The Gaussian integral in the two open string function then becomes
\bea
\int dx^K_idx^L_i e^{iW^\prime}&\to&\left({2\pi\over T_0}\right)^M
\sqrt{JKL\over N+1}\
{\det}^{-1/2}\ \pmatrix{A&B\cr B&A}
\eea
and the prefactors in the expression for the two open string function
become in the limit $J,L\to\infty$,
\bea
{\cal D}_M^{open}(J){\cal D}_{M_1}^{open}(K)
{\cal D}_{M_2}^{open}(K)
{\cal D}_M^{open}(L)&\to&{e^{-(J+L)\sum_m\lambda_m^o/2}
\over K\sqrt{JL}}\left({T_0\over2\pi}\right)^{3M/2}
\prod_{m=1}^{M-1}\left[
2\sinh\lambda^o_m\right]\nonumber\\
&&\prod_{m=1}^{M_1-1}\left[{\sinh K\lambda^{o,1}_m
\over\sinh\lambda^{o,1}_m}\right]^{-1/2}
\prod_{m=1}^{M_2-1}\left[{\sinh K\lambda^{o,2}_m
\over\sinh\lambda^{o,2}_m}\right]^{-1/2}
\eea
Combining these results,
dividing by ${\cal D}^{open}(N+1)e^{NB_0}$ and summing over
$M_1,K$ leads to our expression for the ground state energy shift
\bea
-a\Delta P^-&=&\sum_{K=1}^\infty\sum_{M_1=1}^{M-1}
\left[{{e^{K\sum_m\lambda_m^o/2+(K-1)B_0}}\over\sqrt{K}}
\prod_{m=1}^{M-1}\left[
2\sinh\lambda^o_m\right]^{1/2}\right]^{D-2}\nonumber\\
&&\left[\prod_{m=1}^{M_1-1}\left[{\sinh K\lambda^{o,1}_m
\over\sinh\lambda^{o,1}_m}\right]
\prod_{m=1}^{M_2-1}\left[{\sinh K\lambda^{o,2}_m
\over\sinh\lambda^{o,2}_m}\right]
{\det}\ \pmatrix{A&B\cr B&A}\right]^{-(D-2)/2}
\eea
Doing the three products $\prod_{m=1}^{M-1}\left[
2\sinh\lambda^o_m\right]$ explicitly yields
\bea
-a\Delta P^-&=&\sum_{K=1}^\infty\sum_{M_1=1}^{M-1}
\left[{{e^{K\sum_m\lambda_m^o/2+(K-1)B_0}}\over\sqrt{K}}\right]^{D-2}
\prod_{m=1}^{M-1}\left[
2\sinh\lambda^o_m\right]^{D-2}\nonumber\\
&&\left[{M\over M_1M_2}{\sinh2\sinh^{-1}1\ \sinh2M\sinh^{-1}1\over
\sinh2M_1\sinh^{-1}1\ \sinh2M_2\sinh^{-1}1}
\right]^{-(D-2)/4}\nonumber\\
&&\left[\prod_{m=1}^{M_1-1}\left[2\sinh K\lambda^{o,1}_m
\right]
\prod_{m=1}^{M_2-1}\left[2\sinh K\lambda^{o,2}_m
\right]
{\det}\ \pmatrix{A&B\cr B&A}\right]^{-(D-2)/2}
\eea
\subsection{Correction to the open string gluon energy}
Examination of the open string propagator shows that the gluon
state, the lightest spin one state with energy $\lambda^o_1$
above the ground state,
contributes via the first order term in the expansion of
\bea
\exp\left[T_0{{\bfs q}_{1,f}^o\cdot{\bfs q}_{1,f\i}^o
\sinh\lambda_1^o\over\sinh(N+1)\lambda_1^o}\right]&\sim&
1+T_0{\bfs q}_{1,f}^o\cdot{\bfs q}_{1,i}^o
2\sinh\lambda_1^o e^{-(N+1)\lambda_1^o}
\eea
So to extract the one loop correction we isolate this term
from the two external line propagators in the expression
for the one loop correction to the two point function.
It is then safe to take the $J,L\to \infty$ limit of 
what multiplies these factors. Then in parallel with our
extraction of the correction to the graviton self-energy,
we find
\bea
&&-a\Delta P_{gluon}^-\delta_{kl}=2T_0\sinh\lambda_1^o\sum_{K=1}^\infty\sum_{M_1=1}^{M-1}
\nonumber\\&&\hskip0.75in
e^{K\lambda_1^o}
\VEV{{q}_{1,L}^{ok}{q}_{1,K}^{ol}}
\left[{{e^{K\sum_m\lambda_m^o/2+(K-1)B_0}}\over\sqrt{K}}
\prod_{m=1}^{M-1}\left[
2\sinh\lambda^o_m\right]^{1/2}\right]^{D-2}
\nonumber\\&&\hskip0.75in
\left[\prod_{m=1}^{M_1-1}\left[{\sinh K\lambda^{o,1}_m
\over\sinh\lambda^{o,1}_m}\right]^{-1/2}
\prod_{m=1}^{M_2-1}\left[{\sinh K\lambda^{o,2}_m
\over\sinh\lambda^{o,2}_m}\right]^{-1/2}
{\det}^{-1/2}\ \pmatrix{A&B\cr B&A}\right]^{D-2}
\eea
where the correlator is given by
\bea
\VEV{{q}_{L,1}^{ok}{q}_{K,1}^{ol}}&=&{\int dq_{L,m}^odq_{K,m}^o
{q}_{L,1}^{ok}{q}_{K,1}^{ol}e^{iW^\prime}\over\int dq_{L,m}^odq_{K,m}^o
e^{iW^\prime}}
\eea
Again with the notation
\bea
\pmatrix{A&B\cr B&A}^{-1}&=&\pmatrix{A^\prime&B^\prime\cr B^\prime&A^\prime}
\eea
it follows that
\bea
\VEV{{q}_{L,1}^{ok}{q}_{K,1}^{ol}}&=&\delta_{kl}{B^\prime_{11}\over
T_0}
\eea

\section{Normal Modes}
\label{normalmodes}
A string with $P^+=MaT_0$ is described at a fixed time by
$M$ coordinates $x_i$ or $y_i$, $i=1,\ldots M$. In this article we require
several normal mode decompositions depending on the boundary conditions.
\vskip12pt
\noindent Neumann Open String
\bea
x_i&=&{1\over\sqrt{M}}q_{0}+\sqrt{2\over M}\sum_{m=1}^{M-1}q_{om}
\cos{m\pi(i-1/2)\over M}\\
q_0&=&\sqrt{1\over M}\sum_{i=1}^M x_i,\qquad\quad q_{om}=\sqrt{2\over M}\sum_i x_i\cos{m\pi(i-1/2)\over M}
\label{normalNopen}
\eea
\vskip12pt
\noindent Closed String (Neumann)
\bea
&&\hskip-1inM~{\rm odd}:\nonumber\\
x_i&=&{1\over\sqrt{M}}q_{0}+\sqrt{2\over M}\sum_{m=1}^{(M-1)/2}\left[q_{cm}
\cos{2m\pi(i-1/2)\over M}+q_{sm}\sin{2m\pi(i-1/2)\over M}\right]\\
&&\hskip-1inM~{\rm even}:\nonumber\\
x_i&=&{1\over\sqrt{M}}(q_{0}+q_{sM/2}(-)^i)
\nonumber\\&&
+\sqrt{2\over M}\sum_{m=1}^{M/2-1}\left[q_{cm}
\cos{2m\pi(i-1/2)\over M}+q_{sm}\sin{2m\pi(i-1/2)\over M}\right]\\
q_{cm}&=&\sqrt{2\over M}\sum_i x_i
\cos{2m\pi(i-1/2)\over M},\qquad q_{sm}=\sqrt{2\over M}\sum_i x_i
\sin{2m\pi(i-1/2)\over M}\\
q_{sM/2}&=&\sqrt{1\over M}\sum_{i=1}^M (-)^ix_i,
\qquad {\rm for}\quad M\quad{\rm even},\qquad
q_0=\sqrt{1\over M}\sum_{i=1}^M x_i
\label{normalNclosed}
\eea
\vskip12pt
\noindent Dirichlet Open String
\bea
y_k&=&\sqrt{2\over M}\sum_{m=1}^{M-1}q_{Dm}
\sin{m\pi k\over M}\quad{\rm for}\quad k=1,\ldots,M-1,\qquad
y_M=q_{DM}\\
q_{Dm}&=&\sqrt{2\over M}\sum_{k=1}^{M-1}y_k\sin{m\pi k\over M},\quad
0<m<M,\qquad
q_{DM}=y_M
\label{normalDopen}
\eea
\vskip12pt
\noindent Closed String (Dirichlet)
\bea
&&\hskip-1inM~{\rm odd}:\nonumber\\
y_i&=&{1\over\sqrt{M}}q_{0}+\sqrt{2\over M}\sum_{m=1}^{(M-1)/2}\left[q_{cm}
\cos{2m\pi i\over M}+q_{sm}\sin{2m\pi i\over M}\right]\\
&&\hskip-1inM~{\rm even}:\nonumber\\
y_i&=&{1\over\sqrt{M}}(q_{0}+q_{cM/2}(-)^i)
+\sqrt{2\over M}\sum_{m=1}^{M/2-1}\left[q_{cm}
\cos{2m\pi i\over M}+q_{sm}\sin{2m\pi i\over M}\right]\\
q_{cm}&=&\sqrt{2\over M}\sum_i y_i
\cos{2m\pi i\over M},\qquad q_{sm}=\sqrt{2\over M}\sum_i y_i
\sin{2m\pi i\over M}\\
q_0&=&\sqrt{1\over M}\sum_{i=1}^M y_i,
\qquad\qquad q_{cM/2}=\sqrt{1\over M}\sum_{i=1}^M (-)^iy_i,
\quad({\rm for}~M~{\rm even})
\label{normalDclosed}
\eea
\section{Propagators}
\label{propagators}
\subsection{Neumann open string propagator}
\bea
\VEV{N+1,x^f|0,x^i}^{open}&=&{\cal D}^{open}(N+1)e^{iW_{open}}
\eea
where
\bea
{\cal D}^{\rm open}(N+1)&=&{1\over\sqrt{N+1}}\left({T_0\over2\pi}\right)^{M/2}
\prod_{m=1}^{M-1}\left[{\sinh(N+1)\lambda^o_m
\over\sinh\lambda^o_m}\right]^{-1/2}\\
iW_{open}&=&-{T_0\over2}\bigg[{(q_{0,f}-q_{0,i})^2\over N+1}
+\sum_{m=1}^{M-1}\sinh\lambda^o_m
\bigg((q_{m,i}^2+q_{m,f}^2)\coth(N+1)\lambda^o_m\nonumber\\
&&\hskip2.8in-2{q_{m,i}q_{m,f}\over\sinh(N+1)\lambda^o_m}\bigg)\bigg]\\
\lambda^o_0&=&0,\qquad
\lambda^o_m=2\sinh^{-1}\sin{m\pi\over2M},\quad m=1,\ldots, M-1
\eea
Where the $q_m$'s are the normal mode coordinates for the $x$'s.
The right side is the result of doing the integrations over
all the $x_i^j$ with $i=1,\ldots, M$ and $j=1,\ldots N$. The
propagator spans $N+1$ time steps and this result corresponds to
assigning half the potential energy $T_0\sum_{i=1}^{M-1}(x_{i+1}^j
-x_i^j)^2/2$ to time $j=0$ and half to $j=N+1$.
\subsection{Dirichlet open string propagator}
The Dirichlet open string propagator over a time of $K=N+1$ steps is
evaluated to be
\bea
\VEV{q^f,N+1|q^i,0}^{D}&=&{\cal D}^D(N+1)e^{iW^D}
\eea
where
\bea
{\cal D}^D(N+1)&=&\left({T_0\over2\pi}\right)^{M/2}
\prod_{m=1}^{M}\left[{\sinh(N+1)\lambda^D_m
\over\sinh\lambda^D_m}\right]^{-1/2}\\
 iW^D&=&-{T_0\over2}\sum_{m=1}^{M}\left((q_{Dm}^{f2}+q_{Dm}^{i2})
\sinh\lambda_m^D\coth K\lambda_m^D
-2q_{Dm}^{f}q_{Dm}^{i}{\sinh\lambda_m^D\over
\sinh K\lambda_m^D}\right)\\
 \lambda^D_{m}
&=&\lambda^o_m,\quad m=1,\ldots, M-1,\qquad \lambda^D_{M}=2\sinh^{-1}\sqrt{\kappa\over2}
\eea
We recall that the above expressions give the result of integrating over
all the variables $y^j_i$, for $j=1,\ldots,N$, with half the potential
energy assigned to $j=0,N+1$, which is consistent with the
closure requirement.
\section{Overlap Formulas}
\label{overlap}
\vskip12pt
\noindent{\bf Open-2 Open, Neumann}
\bea
q_0^{(1)}&=&\sqrt{M_1\over M}\ q_0+\sqrt{2\over MM_1}\sum_{m^\prime=1}^{M-1}
 q_{m^\prime}U^{(1)}_{m^\prime 0},\qquad q_m^{(1)}={2\over\sqrt{MM_1}}\sum_{m^\prime=1}^{M-1}
q_{m^\prime}U^{(1)}_{m^\prime m}\\
q_0^{(2)}&=&\sqrt{M_2\over M}\ q_0+\sqrt{2\over MM_2}\sum_{m^\prime=1}^{M-1}
 q_{m^\prime}U^{(2)}_{m^\prime 0},\qquad q_m^{(2)}={2\over\sqrt{MM_2}}\sum_{m^\prime=1}^{M-1}
q_{m^\prime}U^{(2)}_{m^\prime m}\\
U^{(1)}_{m^\prime m}&=&\sum_{i=1}^{M_1}\cos{m^\prime\pi\over M}\left(
i-{1\over2}\right)\cos{m\pi\over M_1}\left(
i-{1\over2}\right)\nonumber\\
&=&{(-)^m\over2}
{\sin(m^\prime\pi M_1/M)\sin(m^\prime\pi/2M)\cos(m\pi/2M_1)\over
\sin^2(m^\prime\pi/2M)-\sin^2(m\pi/2M_1)}\\
U^{(2)}_{m^\prime m}&=&\sum_{i=1+M_1}^{M}\cos{m^\prime\pi\over M}\left(
i-{1\over2}\right)\cos{m\pi\over M_2}\left(
i-M_1-{1\over2}\right)\nonumber\\
&=&-{1\over2}
{\sin(m^\prime\pi M_1/M)\sin(m^\prime\pi/2M)\cos(m\pi/2M_2)\over
\sin^2(m^\prime\pi/2M)-\sin^2(m\pi/2M_2)}
\eea
and we note the identity $q_0^{(1)}\sqrt{M_1}+q_0^{(2)}\sqrt{M_2}
=q_0\sqrt{M}$, as expected from the fact that $q_0/\sqrt{M}$ is the center
of momentum of the open string.

We can also express the $q$'s in terms of the $q^{(1)},q^{(2)}$'s:
\bea
q_0&=&q_0^{(1)}\sqrt{M_1\over M}+q_0^{(2)}\sqrt{M_2\over M}\\
q_{m^\prime}
&=&\sqrt{2\over MM_1}\left(q_0^{(1)}U^{(1)}_{m^\prime0}
+\sqrt{2}\sum_{m=1}^{M_1-1}q_m^{(1)}U^{(1)}_{m^\prime m}\right)
\nonumber\\&&\qquad
+\sqrt{2\over MM_2}\left(q_0^{(2)}U^{(2)}_{m^\prime0}+\sqrt{2}
\sum_{m=1}^{M_2-1}q_m^{(2)}U^{(2)}_{m^\prime m}\right)
\eea
\vskip12pt
\noindent{\bf Open-2 Open, Dirichlet}
\bea
q_{DM_1}^{(1)}&=&\sqrt{2\over M}\sum_{m^\prime=1}^{M-1}
 q_{Dm^\prime}\sin{m^\prime\pi M_1\over M},\qquad
q_{Dm}^{(1)}={2\over\sqrt{MM_1}}\sum_{m^\prime=1}^{M-1}
q_{Dm^\prime}U^{D(1)}_{m^\prime m}\\
q_{DM_2}^{(2)}&=&y_M=q_{DM},\qquad
q_{Dm}^{(2)}={2\over\sqrt{MM_2}}\sum_{m^\prime=1}^{M-1}
q_{Dm^\prime}U^{D(2)}_{m^\prime m}\\
U^{D(1)}_{m^\prime m}&=&\sum_{k=1}^{M_1-1}\sin{m^\prime\pi k\over M}
\sin{m\pi k\over M_1}={(-)^m\over4}{\sin(m\pi/M_1)\sin(m^\prime\pi M_1/M)
\over \sin^2(m^\prime\pi/2M)-\sin^2(m\pi/2M_1)}\nonumber\\
U^{D(2)}_{m^\prime m}&=&\sum_{i=1+M_1}^{M-1}\sin{m^\prime\pi k\over M}
\sin{m\pi (k-M_1)\over M_1}\nonumber\\
&=&{(-)^{m^\prime}\over4}{\sin(m\pi/(M-M_1))\sin(m^\prime\pi(M-M_1)/M)
\over \sin^2(m^\prime\pi/2M)-\sin^2(m\pi/2(M-M_1))}\eea
\vskip12pt
\noindent{\bf 3 Zero-momentum Tachyon Vertex}
\bea
V_3&=&{1\over\sqrt{|P^+_1P^+_2P_3^+|}}
\left|{P^+_1\over P^+_3}\right|^{(P_1^{+2}+P_2^{+2}+P_1^+P_2^+)/
P_2^+P_3^+}
\left|{P^+_2\over P^+_3}\right|^{(P_1^{+2}+P_2^{+2}+P_1^+P_2^+)/
P_1^+P_3^+}\\
P_3^+&=&-P_1^+-P_2^+
\eea


\end{document}